\documentclass[preprint,authoryear,11pt,letterpaper]{elsarticle}
\usepackage[top=0.75in, bottom=0.75in, left=0.75in, right=0.75in]{geometry}

\usepackage{chngcntr}
\usepackage{amsmath,amsthm}
\usepackage{amssymb}
\usepackage{bbm}
\usepackage{float}
\usepackage{amsfonts}
\usepackage{graphicx}
\usepackage{subfigure}
\graphicspath{{Graphics/}}
\usepackage[ruled,vlined,algo2e]{algorithm2e}

\usepackage{algorithm}
\usepackage{algorithmic}
\usepackage{tabularx,booktabs}
\usepackage{float}
\usepackage{graphicx,color}
\usepackage[T1]{fontenc}

\usepackage{xcolor}
\definecolor{darkblue}{rgb}{0.0,0.5,0.5}
\definecolor{blue}{rgb}{0.0,0.0,1}
\usepackage[colorlinks]{hyperref}
\hypersetup{colorlinks,breaklinks,linkcolor=blue,urlcolor=blue,anchorcolor=blue,citecolor=blue}
\usepackage{booktabs,caption}
\usepackage{multirow}
\usepackage{lscape}
\usepackage{epstopdf}
\usepackage{lineno}

\usepackage{microtype}
\usepackage{lineno}
\journal{Arxiv}

\begin{document}
\begin{frontmatter}

\title{Probabilistic forecasting of bus travel time with a Bayesian Gaussian mixture model}

\author[1,4]{Xiaoxu Chen}
\ead{xiaoxu.chen@mail.mcgill.ca}
\author[1,4]{Zhanhong Cheng}
\ead{zhanhong.cheng@mail.mcgill.ca}%
\author[2]{Jian Gang Jin}
\ead{jiangang.jin@sjtu.edu.cn}
\author[3,4]{Martin Tr\'epanier}
\ead{mtrepanier@oolymtl.ca}
\author[1,4]{Lijun Sun\corref{cor1}}
\ead{lijun.sun@mcgill.ca}

\cortext[cor1]{Corresponding author.}
\address[1]{Department of Civil Engineering, McGill University, Montreal, QC H3A 0C3, Canada}
\address[2]{School of Naval Architecture, Ocean \& Civil Engineering, and State Key Laboratory of Ocean Engineering,\\ Shanghai Jiao Tong University, Shanghai, 200240, China}
\address[3]{Department of Mathematics and Industrial Engineering, Polytechnique Montreal, Montreal, QC H3T 1J4, Canada}
\address[4]{Interuniversity Research Centre on Enterprise Networks, Logistics and Transportation (CIRRELT)}

\begin{abstract}
Accurate forecasting of bus travel time and its uncertainty is critical to service quality and operation of transit systems; for example, it can help passengers make better decisions on departure time, route choice, and even transport mode choice and also support transit operators to make informed decisions on tasks such as crew/vehicle scheduling and timetabling. However, most existing approaches in bus travel time forecasting are based on deterministic models that provide only point estimation. To this end, we develop in this paper a Bayesian probabilistic forecasting model for bus travel time. To characterize the strong dependencies/interactions between consecutive buses, we concatenate the link travel time vectors and the headway vector from a pair of two adjacent buses as a new augmented variable and model it with a constrained Multivariate Gaussian mixture distributions. This approach can naturally capture the interactions between adjacent buses (e.g., correlated speed and smooth variation of headway), handle missing values in data, and depict the multimodality in bus travel time distributions. Next, we assume different periods in a day share the same set of Gaussian components but different mixing coefficients to characterize the systematic temporal variations in bus operation. For model inference, we develop an efficient Markov chain Monte Carlo (MCMC) sampling algorithm to obtain the posterior distributions of model parameters and make probabilistic forecasting. We test the proposed model using the data from a twenty-link bus route in Guangzhou, China. Results show our approach significantly outperforms baseline models that overlook bus-to-bus interactions in terms of both predictive means and distributions. Besides forecasting, the parameters of the proposed model contain rich information for understanding/improving the bus service, e.g., analyzing link travel time and headway correlation using correlation matrices and understanding time-varying patterns of bus fleet operation from the mixing coefficients.
\end{abstract}


\end{frontmatter}
\section{Introduction}
Cities are now facing severe traffic congestion and air pollution due to the over-reliance on cars. Promoting public transportation is one of the most effective and strategic ways to achieve urban sustainability. Although public transport agencies have developed bus systems for many decades, many people would not like to take the bus for various reasons, such as vulnerable reliability of travel time, uncomfortable riding environment, and stations far away from home. Survey studies have shown that passengers highly care about the bus travel/arrival time forecasting and its reliability (uncertainty) \citep{lam2001value}. A probabilistic forecasting of bus travel time provides both the expected value and the uncertainty, which not only helps bus agencies to design robust bus management strategies (e.g., bus timetable, bus priority signal control, bus bunching control) to improve bus services, but also aids bus passengers to make better travel plans regarding departure time, route choice, and even transport mode choice \citep{cats2017modeling}.

Existing studies on bus travel time forecasting mainly center on predicting travel time values (deterministic forecasting) but ignore the importance of travel time uncertainty \citep{ricard2022predicting}. There are many deterministic bus travel time forecasting methods such as historical average (HA) \citep{farhan2002bus}, Autoregressive Integrated Moving Average (ARIMA) \citep{madzlan2010arima}, Artificial Neural Network (ANN) \citep{chien2002dynamic,gurmu2014artificial}, Support Vector Machine (SVM) \citep{bin2006bus,yu2011bus}, Kalman Filtering (KF) \citep{cathey2003prescription}, Long Short-Term Memory (LSTM) \citep{osman2021application,alam2021predicting}, and various hybrid models \citep{yu2018prediction,zhang2021automatic}, to name just a few. A major limitation of deterministic models is that they cannot give the uncertainty of the forecasting.
Therefore, probabilistic forecasting (i.e., forecast the distribution of bus travel time) is favored over deterministic forecasting \citep{yetiskul2012public}. A critical step in probabilistic bus travel time forecasting is constructing an appropriate probabilistic distribution for bus travel time, which is very challenging because of the following difficulties: 1) complex correlations between link travel times within a bus route, 2) interactions between adjacent buses, and 3) bus travel time distributions are usually not normal and exhibit long-tailed and multimodal characteristics \citep{ma2016modeling}. A recent study \citep{chen2022bayesian} demonstrated that link travel times in a bus route exist complex local and long-range correlations, and bus travel times on different links are often positively correlated because of factors like bus bunching. However, the very limited existing research on probabilistic bus travel time forecasting often failed to consider the complex link travel time correlation. For example, \citet{huang2021bus} and \citet{ricard2022predicting} did not model the link travel time correlation; \citet{ma2017estimation} and \citet{buchel2022modeling} only considered the (local) correlation between adjacent link travel times. Moreover, only a few works considered the interactions between buses in travel time forecasting \citep[e.g.,][]{dai2019bus}, the potential of leveraging the information from neighboring buses to improve the forecasting is still huge. Lastly, most research applied a unimodal Gaussian assumption \citep{dai2019bus,chen2022bayesian}, which failed to realistically capture the bus travel time in reality.

To address the above issues, this paper develops a Bayesian probabilistic model for bus travel/arrival time forecasting. Specifically, we concatenate the link travel time vectors and the headway vector of each two adjacent buses as a new augmented random variable. By utilizing the inherent relationship (linear equality constraints) between link travel times and headways, our approach naturally captures the interactions between adjacent buses and handles missing values in data. To capture the multimodality of bus travel time distribution, we model the augmented random variable with multivariate Gaussian mixture distributions truncated by the hyperplane defined by the headway constraints. Moreover, we borrow the idea of the topic model \citep{blei2003latent} to capture temporal differences in bus travel time: different periods in a day share the same set of Gaussian components but different mixing coefficients. Next, we develop an efficient Markov chain Monte Carlo (MCMC) sampling algorithm to obtain the posterior distribution of model parameters. Based on the estimated probabilistic model, we could make conditional probabilistic forecasting for bus travel time in an autoregressive way (the forecasting of a bus relies on the forecasting of the previous bus). We test the proposed probabilistic forecasting model using a dataset from a twenty-link bus route in Guangzhou, China. Results show our approach that considers the dependencies between adjacent buses and the headway relationships significantly outperforms baseline models that do not consider these factors, in terms of both predictive means and distributions. Besides forecasting, the parameters of the proposed model contain rich information for understanding/improving the bus service, e.g., analyzing link travel time correlation using correlation matrices and understanding temporal patterns of the bus route from mixing coefficients.





The contributions of this work include three aspects. First, we design a new representation for link travel times of a bus route by an augmented random variable. This new representation naturally handles various difficulties in modeling bus travel time, including correlations between link travel time in a bus route, interactions between adjacent buses, and missing values in data. Second, we integrate Gaussian mixture models into a topic model framework as the distribution for the augmented random variable. The Gaussian mixture models can depict the multimodality in bus travel time distributions, and the topic model can reflect different mixing patterns in bus travel time at different periods of a day. Third, we develop a Gibbs sampling algorithm to obtain the posterior distributions of model parameters and enable probabilistic forecasting. Experiments in a real-world dataset show the proposed model can accurately predict the bus travel time distributions and discover patterns in link travel time correlations.

The remainder of this paper is organized as follows. In Section~\ref{relatedwork}, we review previous studies on bus travel time forecasting. In Section~\ref{sec:method}, we describe the problem and present the Bayesian Gaussian
mixture model. Next, in Section~\ref{casestudy}, we demonstrate the capability of the proposed model by experiments using real-world data. Finally, we conclude our study, summarize our main findings, and discuss future research directions in Section~\ref{conclusion}.

\section{Related Work}
\label{relatedwork}
There have been some bus travel time forecasting models. However, most of them concentrate on forecasting the travel time values, instead of forecasting the travel time distribution. In this literature review, we categorize bus travel time forecasting models into deterministic forecasting model and probabilistic forecasting model. After the introduction of deterministic forecasting model, we pay more attention on the studies related to probabilistic forecasting model for bus travel time.

\subsection{Deterministic Forecasting Model}
Deterministic forecasting model for bus travel time includes time series models and machine learning approaches. Time series model often considers bus travel time as a function of its past observations. Using time series model to forecast travel time, one important step is to construct a standard time series of bus travel time from bus data. \citet{farhan2002bus} used HA model to predict the bus travel time. \citet{madzlan2010arima} applied ARIMA model to forecast bus route travel time. The forecasting accuracy of time series model highly depend on the characteristics of historical travel time data. If a time series has a stationary pattern or mixture of patterns, time series model can perform well. However, bus travel time is not a standard stationary time series due to the factors like road congestion, dynamic passenger flow, traffic accidents, etc. Therefore, time series model is strictly limited for bus travel time forecasting.

Machine learning approaches have the ability to reveal complex patterns and learn non-linear relationships from data. Existing machine learning models for bus travel time forecasting include ANN, SVM, KF, and the hybrid models. \citet{chien2002dynamic} developed two ANN-based models to forecast bus arrival time under link-based and stop-based route constructions; results show that these two ANN-based models have good performance. Similarly, the studies by \citet{gurmu2014artificial}, \citet{jeong2004bus} indicate that ANN-based models outperform HA and regression models for bus travel time forecasting. SVM has been applied as a useful method for forecasting bus travel time. Some studies \citep{bin2006bus,yu2011bus,yang2016bus,xu2017bus,MA2019536} have shown SVM outperforms linear regression and time series models. KF is an efficient learning algorithm as it has the ability to update the time-dependent state when new observations become available continuously. \citet{cathey2003prescription} proposed KF model for bus travel time forecasting using real-time and historical data. However, this work did not compared KF model with other developed machine learning models. Some hybrid methods are developed to combine the advantages of these machine learning models. \citet{yu2018prediction} proposed an approach combining k-nearest neighbors algorithm (KNN) and Random Forest (RF); results shows that the hybrid model outperformed KNN, SVM, RF; however, this method has the problem of low computational efficiency. \citet{zhang2021automatic} developed a hybrid method to combine SVM with KF, RF and ARIMA respectively; results show that SVM-KF outperforms other hybrid models. \citet{chen2004dynamic} combined ANN and KF algorithm to forecast the arrival time. Based on this research, \citet{bai2015dynamic} replaced ANN with SVM, and results show that SVM-KF performs better than ANN-KF. \citet{kumar2018hybrid} used KNN algorithm to identify significant input variables, and then combining exponential smoothing technique with recursive estimation scheme based on Kalman Filtering method to forecast bus travel time.
Recently, \citet{osman2021application} and \citet{alam2021predicting} used LSTM to forecast bus travel timne; results show that LSTM performs better than ANN, SVR, ARIMA and HA. Similarly, \citet{he2020learning} applied LSTM for bus travel time prediction but they took consideration into the impact of heterogeneous traffic patterns. \citet{petersen2019multi} proposed a deep neural network with the combination of convolutional and LSTM layers for bus travel time prediction.

\subsection{Probabilistic Forecasting Model}
There are only a few studies on  probabilistic forecasting for bus travel time. \citet{dai2019bus} proposed a probabilistic model to estimate bus travel time considering the link running time and station dwell time. The authors assumed that: the link travel time is composed of the link running time and the station dwell time, and they are independent; the link running time follows shifted log-normal distribution; the station dwell time is the sum of the queueing time, the passengers loading/unloadding time, and the merging time (the bus merges into the main road traffic). They did not model the correlations between link running times and station dwell times while the correlations are very important for bus travel time forecasting. \citet{ma2017estimation} proposed a generalized Markov chain approach for estimating the probability distribution of bus trip travel times from link travel time distributions and takes into consideration correlations in time and space. They first used clustering method to cluster link travel time observations, and then using a logit model to develop the transition probability estimation; finally, using a Markov chain procedure, the probability distribution of the trip travel time can be estimated. However, the Markov chain in this framework only models the correlation between adjacent link travel times but ignores the long-range correlation. \citet{huang2021bus} proposed two data-driven methods based on the Functional Data Analysis (FDA) and Bayesian Support Vector Regression (BSVR) to forecast the distribution of bus travel time. Both FDA and BSVR are essentially kernel methods. FDA approach is a well-proven mathematical way to describe the stochastic process of link travel time and can provide a continuous-time link travel time forecasting, while BSVR can provide a discrete-time link travel time forecasting with a prescribed discretization interval. The authors utilize the FDA and BSVR for each specific link and then add up relevant link travel times; thus they ignore the link travel time correlation. \citet{buchel2022modeling} proposed a hidden Markov chain framework to estimate bus travel time distribution. This framework can capture the dependency structure of consecutive link running times and includes conditional correlations. Moreover, it also captures the dependency of consecutive link dwell times. However, this model only describe the correlations between consecutive link travel times and fails to model long-range correlations. \citet{ricard2022predicting} proposed two types of probabilistic models: similarity-based density estimation models and a smoothed Logistic Regression for probabilistic classification model. Similarity-based density estimation models first find the set of similar trips and then estimate the density of the particular set by fitting a parametric, semi-parametric or non-parametric model. Multinomial logistic regression is used for probabilistic classification and its generated probability mass function can be smoothed into a probability density function. The authors developed these two methods in order to make a long-term forecast for bus travel time, thus they did not consider to use the feature of link travel time. \citet{buchel2022we} proposed Bayesian network to forecast bus travel time distribution. They assumed that: the dwell time of a given bus at a given stop depends on the dwell time of the same bus at the previous stop, dwell time of the previous bus at the same stop, and the headway from the previous bus; the running time depends on the running time of the same bus in the previous link and the running time of the previous bus in the same link. This Bayesian framework only consider the dependency between adjacent link travel times but ignore the long-range correlations in bus link travel times.

Although work on probabilistic forecasting for bus travel time is scare, there are some studies on modeling bus travel time distributions, mostly with the objective of quantifying the bus travel time reliability \citep{buchel2020review}. Bus travel time distribution modeling can provide a foundation for probabilistic forecasting. \citet{taylor1982travel} collected bus data and pointed out bus link travel time follows a normal distribution. \citet{mazloumi2010using} explored the travel time distributions for different departure time windows at different times of the day; results show that in narrower departure time windows, bus travel time distributions are best characterized by normal distributions. For wider departure time windows, peak-hour travel times follow normal distributions, while off-peak travel times follow log-normal distributions. Similarly, \citet{rahman2018analysis} analyzed the bus travel time distributions for different spatial horizon; results show that log-normal distribution is more appropriate for short-term horizon, while normal distribution is more suitable for long-term horizon. \citet{uno2009using} reveled bus link travel time on arterial roadway is positively skewed and generally follows a log-normal distribution. \citet{kieu2015public} also recommended a log-normal distribution as the best fit of bus travel time on urban roads. \citet{buchel2018modelling} compared the unimodal distributions including normal, weibull, log-normal, gamma, cauchy, and logistic distribution; results show that the log-normal probability distribution is a good fit for bus travel times at peak and off-peak conditions. \citet{chepuri2020development}  compared the Burr, generalized extreme value (GEV), and Log-normal distributions for bus travel time, and results shows that GEV distribution performs the best. Similarly, \citet{harsha2021probability} considered seven travel time distributions (including Burr, GEV, Gamma, log-logistic, log-normal, normal and Weibull distributions) and evaluated their performance; results show that GEV distribution performs the best. \citet{ma2016modeling} compared unimodal distribution and multi-modal distribution for bus link travel time. They found that the normal, log-normal, logistic, log-logistic, and Gamma distributions have a relatively similar performance, and the Gaussian
mixture model performs much better in terms of accuracy, robustness, and interpretability.

In summary, existing studies mainly concerned with the deterministic forecasting instead of probabilistic forecasting for bus travel time. Although a few studies have developed probabilistic forecasting for bus travel time, there are some limitations: (1) ignoring the complex link travel time correlation/covariance; (2) ignoring the correlation between buses.

\section{Methodology}
\label{sec:method}
\subsection{Problem Description}
A bus link (or simply a link) is the directional road segment between two adjacent bus stops in a bus route. In this paper, the bus travel time on the $j$-th link is defined as the time difference between the arrival of a bus at the $j$-th and the $(j+1)$-th bus stop, including the bus waiting time at the $j$-th bus stop. Link travel time of buses can be obtained from various types of data sources, such as smart card data, automatic vehicle location (AVL) data, and automatic bus announcing system. Next, we denote $\ell_{i, j}$ to be the link travel time of the $i$-th bus on the $j$-th link. With these definitions, the trip travel time of the $i$-th bus from stop $j_1$ to stop $j_2$ can be readily calculated by $\sum_{j=j_1}^{j_2} \ell_{i,j}$.

This paper focuses on forecasting the travel time of a bus on its upcoming links and trips. A previous work \citep{chen2022bayesian} has shown that link travel time within a single bus trip can be significantly correlated, and using such correlation can improve bus travel time forecasting. A limitation of the previous work is that the dependencies of the travel time among different buses are ignored. Considering the close spatiotemporal distance and similar traffic conditions of two adjacent buses, it is tempting to use the travel time information of a leading bus to forecast the travel time of the next (following) bus. Inspired by this, on top of the previous work, this paper further leverage the travel time correlation between adjacent buses to improve bus travel time forecasting.

\subsection{Augmented Random Variable}
The link travel time of bus $i$ on a bus route with $n$ links ($n+1$ bus stops) can be aggregated into a vector $\boldsymbol{\ell}_i=\left[\ell_{i,1},\ell_{i,2},\cdots,\ell_{i,n}\right]^\top$. We define an augmented random variable $\boldsymbol{x}$ to capture the link travel time correlation between two adjacent buses. The link travel time and the headway of each two adjacent buses (a following bus $i$ and a leading bus $i-1$) produce a sample of $\boldsymbol{x}$:
\begin{equation}
    \boldsymbol{x}_{i}=\begin{bmatrix}\boldsymbol{\ell}_{i} \\ \boldsymbol{\ell}_{i-1} \\ \boldsymbol{h}_{i} \end{bmatrix}=\left[\ell_{i,1},\ell_{i,2},\cdots,\ell_{i,n}, \ell_{i-1,1}, \ell_{i-1,2}, \cdots, \ell_{i-1,n}, h_{i,1},h_{i,2},\cdots, h_{i,n}\right]^\top\in\mathbb{R}^{3n},
\end{equation}
where the headway $h_{i,j}$ is the time interval between the arrival of the $i$-th bus and the $(i-1)$-th bus at the $j$-th bus stop (we do not count the headway at the last bus stop). Note that there is an inherent relationship between the link travel time and the headway:
\begin{equation}\label{eq:headway_constraint}
    h_{i, j+1}-h_{i, j}+\ell_{i,j}-\ell_{i-1, j}=0,\,1\leq j \leq n-1.
\end{equation}
Therefore, the $3n$ dimensional random variable $\boldsymbol{x}$ has only $2n+1$ degrees of freedom. The inclusion of headway explicitly bonds $\boldsymbol{\ell}_i$ with $\boldsymbol{\ell}_{i-1}$, which efficiently utilizes the link travel time of the leading bus in forecasting the link travel time of the following bus.

\begin{figure}[htbp]
    \centering
    \subfigure{\includegraphics[width = 0.6\textwidth]{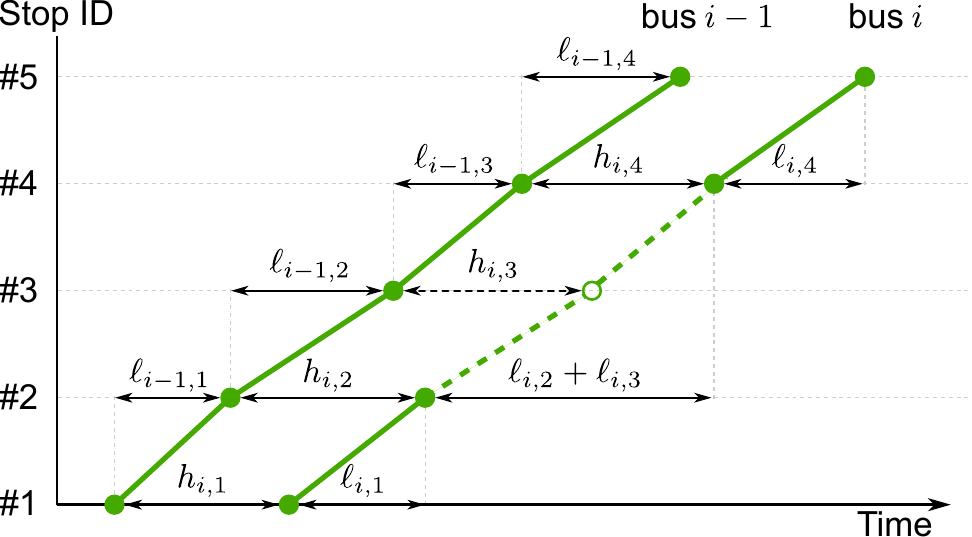}}
    \caption{Trajectories of two adjacent buses. Solid dots: observed arrival time at bus stations. The hollow dot: unknown arrival time at the bus station.} \label{link_headway}
\end{figure}

Missing and ragged values are unavoidable in real-world link travel time data. Using the relationship between adjacent buses can enhance the accuracy of the missing value imputation. We use the method proposed by \citet{chen2022bayesian} to jointly formulate the missing/ragged values and the headway constraints from Eq.~\eqref{eq:headway_constraint}. Consider the example in Fig.~\ref{link_headway} where the arrival time of bus $i$ at stop $\#3$ cannot be observed. The headway $h_{i,3}$ becomes a missing value. Although individual values of $\ell_{i,2}$, $\ell_{i,3}$ are missing, the sum of the link travel time $\left(\ell_{i,2} + \ell_{i,3}\right)$ can be inferred from the bus arrival time at the upstream and the downstream stops, which is a case of ragged value. Next, the missing/ragged values and headway constraints can be summarized into a linear equation:
\begin{equation}\label{eq:linear constraints}
    \begin{matrix}
    \addtocounter{MaxMatrixCols}{13}
    \underbrace{
        \begin{bmatrix}
          1&		0&		0&		0&		0&		0&   0&		0&		0&		0&		0 & 0\\
          0&		1&		1&		0&		0&		0&   0&		0&		0&		0&		0 & 0\\
          0&		0&		0&		1&		0&		0&   0&		0&		0&		0&		0 & 0\\
          0&		0&		0&		0&		1&		0&   0&     0&		0&		0&		0 & 0\\
          0&		0&		0&		0&		0&		1&   0&		0&		0&		0&		0 & 0\\
          0&		0&		0&		0&		0&		0&   1&		0&		0&		0&		0 & 0\\
          0&		0&		0&		0&		0&		0&   0&		1&		0&		0&		0 & 0\\
          0&		0&		0&		0&		0&		0&   0&		0&		1&		0&		0 & 0\\
          1&		0&		0&		0&		-1&		0&   0&		0&		-1&		1&		0 & 0\\
          0&		1&		0&		0&		0&		-1&  0&		0&		0&		-1&		1 & 0\\
          0&		0&		1&		0&		0&		0&   -1&	0&		0&		0&		-1 & 1\\
    \end{bmatrix}}_{\mathbf{G}_{i}}
    \boldsymbol{x}_{i} =
    \begin{bmatrix}
          \ell_{i,1}\\
          \ell_{i,2}+ \ell_{i,3}\\
          \ell_{i,4}\\
          \ell_{i-1,1}\\
          \ell_{i-1,2}\\
          \ell_{i-1,3}\\
          \ell_{i-1,4}\\
          h_{i,1}\\
          h_{i,2}-h_{i,1}+\ell_{i,1}-\ell_{i-1,1}\\
          h_{i,3}-h_{i,2}+\ell_{i,2}-\ell_{i-1,2}\\
          h_{i,4}-h_{i,3}+\ell_{i,3}-\ell_{i-1,3}\\
    \end{bmatrix} =
    \underbrace{\begin{bmatrix}
          r_{i,1}\\
          r_{i,2}\\
          r_{i,3}\\
          r_{i,4}\\
          r_{i,5}\\
          r_{i,6}\\
          r_{i,7}\\
          r_{i,8}\\
          0 \\
          0 \\
          0 \\
    \end{bmatrix}}_{\boldsymbol{r}_{i}} \\
    \end{matrix},
\end{equation}
where we call $\mathbf{G}_i$ the \textit{alignment matrix} and $\boldsymbol{r}_i$ the \textit{recording vector} for bus $i$. An alignment matrix is a matrix with elements in $\left\{-1, 0, 1\right\}$ that encodes missing/ragged positions and headway constraints. The recording vector $\boldsymbol{r}_i$ records all observed information attached with $\boldsymbol{x}_i$. The hyperplane defined by Eq.~\eqref{eq:linear constraints} is the support of random variable $\boldsymbol{x}$.

Alignment matrices and recording vectors can be directly accessed from the source data, but the values of $\boldsymbol{x}_i$ are not always available because of the data missing problem. Next, the main task is to estimate the probability distribution (a Gaussian mixture model) of the augmented random variable $\boldsymbol{x}$ using historical alignment matrices $\left\{ \mathbf{G}_i \right\}$ and recording vectors $\left\{\boldsymbol{r}_i \right\}$. After getting the probabilistic distribution of $\boldsymbol{x}$, we can forecast the bus travel time on upcoming links by calculating the conditional probability given the travel time of the leading buses and upstream links. The forecasting procedure is elucidated in Section~\ref{sec:forecast}.

\subsection{Bayesian Multivariate Gaussian Mixture Model}
The distributions of bus link travel times are often positively skewed, heavy-tailed, and sometimes multi-modal. \citet{ma2016modeling} compared several unimodal distributions (including normal, log-normal, logistic, log-logistic, and Gamma distributions) and multi-modal distributions for bus travel time, and suggested using Gaussian mixture models for bus link travel times. Therefore, we use multivariate Gaussian mixture distributions to model the augmented random variable $\boldsymbol{x}$. Moreover, we divide a day into $T$ periods and assume the mixing coefficients are different for each period to capture the temporal differences. When not considering the headway constraints in Eq.~\eqref{eq:linear constraints}, the augmented random variable at the $t$-th period follows a multivariate Gaussian mixture model:
\begin{equation}\label{Multi-variate Density}
    p^{t}\left(\boldsymbol{x}^t \right) = \sum_{k=1}^{K}\pi_k^{t} \mathcal{N}\left(\boldsymbol{x}^t\mid\boldsymbol{\mu}_k,\boldsymbol{\Sigma}_k\right),
\end{equation}
where we use a superscript $(\cdot)^{t}$ to mark the time period; $K$ is the number of components; $\pi_k^t$ is a mixing coefficient with $\sum_{k=1}^{K}\pi_k^t=1$ and $\pi_k^t \ge 0$; the $k$-th component follows a multivariate Gaussian distribution with a mean vector $\boldsymbol{\mu}_k \in \mathbb{R}^{3n} $ and a covariance matrix $\boldsymbol{\Sigma}_k \in \mathbb{R}^{3n \times 3n}$. When considering the linear constraints in Eq.~\eqref{eq:linear constraints}, the distribution of $\boldsymbol{x}_t$ is the multivariate Gaussian mixture model in Eq.~\eqref{Multi-variate Density} truncated on the hyperplane defined by Eq.~\eqref{eq:linear constraints}.


\begin{figure}[!ht]
    \centering
    \subfigure{
        \centering
        \includegraphics[width = 0.6\textwidth]{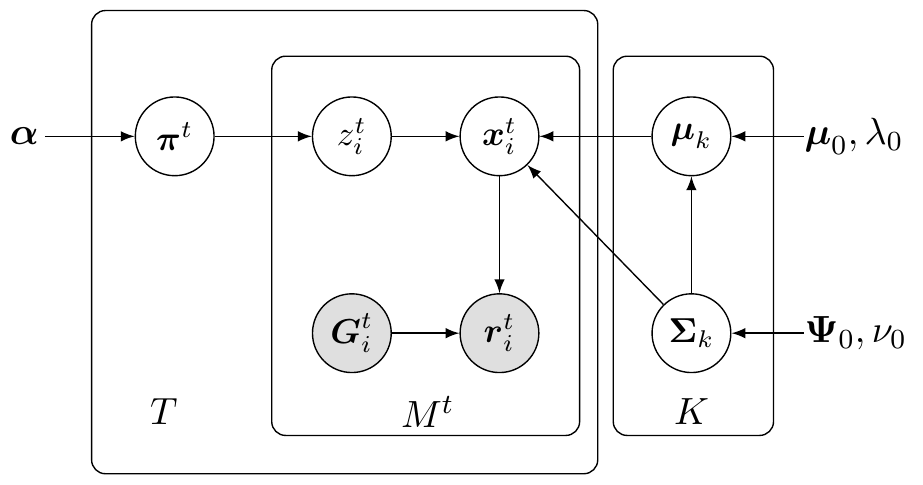}
    }
    \caption{The graphical illustration of Bayesian Gaussian mixture model.}
    \label{fig:graphical model}
\end{figure}

The augmented link travel time of each period is characterized by a mixture of several shared Gaussian distributions; this structure is essentially similar to the classical topic model \citep{blei2003latent}. Fig.~\ref{fig:graphical model} shows the overall graphical representation of our Bayesian multivariate Gaussian mixture model that can handle missing, ragged values, and headway constraints. Assume there are $M^t$ buses at period $t$, we have $\left\{\boldsymbol{x}_i^t \right\}_{t=1, i=1}^{T,M^t}$ to be a set of ``latent'' realizations/samples drawn from the multivariate Gaussian mixture distributions truncated on the hyperplanes from Eq.~\eqref{eq:linear constraints}. We need to estimate parameters $\left\{\boldsymbol{\pi}^t = [\pi_1^t, \pi_2^t, \cdots, \pi_K^t ]^{\top}\right\}_{t=1}^T$, $\left\{\boldsymbol{\mu}_k\right\}_{k=1}^K$, and $\left\{\boldsymbol{\Sigma}_k\right\}_{k=1}^K$ using the alignment matrices $\mathcal{G}=\left\{{\mathbf{G}}_{i}^t \right\}_{t=1,i=1}^{T,M_t}$ and the recording vectors $\mathcal{R} = \left\{\boldsymbol{r}_{i}^t \right\}_{t=1,i=1}^{T,M_t}$. In a Bayesian setting, we use a conjugate Gaussian-inverse-Wishart prior on $\boldsymbol{\mu}_k$ and $\boldsymbol{\Sigma}_k$ and a Dirichlet prior on $\boldsymbol{\pi}^t$ for efficient inference \citep{gelman1995bayesian}. The overall data generation process is summarized as:
\begin{align}
    \boldsymbol{\pi}^t &\sim \text{Dirichlet}\left(\boldsymbol{\alpha}\right),
    \label{eq:pi prior}
    \\
    \boldsymbol{\Sigma}_k& \sim \mathcal{W}^{-1}\left(\boldsymbol{\Psi}_0,\nu_0\right), \label{eq:cov prior} \\
    \boldsymbol{\mu}_k &\sim \mathcal{N}\left(\boldsymbol{\mu}_{0}, \frac{1}{\lambda_{0}}\boldsymbol{\Sigma}_k \right), \label{eq:mean prior} \\
    z_{i}^t &\sim \text{Categorical}\left(\boldsymbol{\pi}^t\right), \label{eq:zpdf}
    \\
    \boldsymbol{x}_{i}^t\mid z_{i}^t =k&\sim \mathcal{N}\left(\boldsymbol{\mu}_{k},\boldsymbol{\Sigma}_{k}\right), \label{xpdf}\\
    \boldsymbol{r}_{i}^t &= \mathbf{G}_{i}^t \boldsymbol{x}_{i}^t, \label{eq:linear}
\end{align}
where $\boldsymbol{\alpha}$ is the concentration parameter of the Dirichlet distribution; $\mathcal{W}^{-1}\left(\boldsymbol{\Psi}_0,\nu_0\right)$ is the inverse-Wishart distribution with a scale matrix $\boldsymbol{\Psi}_0$ and $\nu_0$ degrees of freedom; $\boldsymbol{\mu}_0$ and $\lambda_0$ are parameters for the Gaussian prior.

\subsection{Model Inference}

Based on the graphical model illustrated in Fig.~\ref{fig:graphical model}, we can derive an efficient MCMC scheme using Gibbs sampling. For simplicity, we let $\mathcal{X} = \left\{\boldsymbol{x}_{i}^t \right\}_{t=1,i=1}^{T,M^t}$ denote the full set of the ``latent'' augmented variable for the $\sum_{t=1}^TM^t$ bus pairs, and let $\Theta=\left\{\boldsymbol{\mu}_0,\lambda_0, \boldsymbol{\Psi}_0, \nu_0\right\}$ denote the set of hyperparameters for the Gaussian-inverse-Wishart prior distribution in Eqs.~\eqref{eq:cov prior} and \eqref{eq:mean prior}. Moreover, we let $\mathcal{X}_{k}^{t}$ denote the set of data vectors at period $t$ belonging to mixture component $k$ and $\mathcal{X}_k$ denote the set of all data vectors belonging to mixture $k$. We start the Gibbs sampling with random initialization for all variables and then iteratively sample each variable from its conditional distribution on other variables:

\begin{itemize}
    \item Sample $\boldsymbol{\pi}_t$ from $p\left(\boldsymbol{\pi}^t\mid \boldsymbol{z}^t,\boldsymbol{\alpha}\right)$. The conditional distribution is $p\left(\boldsymbol{\pi}^t\mid \boldsymbol{z}^t,\boldsymbol{\alpha}\right) \propto p\left(\boldsymbol{\pi}^t\mid \boldsymbol{\alpha}\right) p\left(\boldsymbol{z}^t\mid \boldsymbol{\pi}^t\right)$. The prior distribution $p\left(\boldsymbol{\pi}^t\mid \boldsymbol{\alpha}\right) = \text{Dirichlet}\left(\boldsymbol{\pi}^t\mid \boldsymbol{\alpha}\right) \propto \prod_{k=1}^{K}{\pi_{k}^{t}}^{\alpha_k-1}$, and $p\left(\boldsymbol{z}^t\mid \boldsymbol{\pi}^t\right)$ can be seen as a multinomial distribution $p\left(\boldsymbol{z}^t\mid \boldsymbol{\pi}^t\right)=\text{Multinomial}_K\left(\boldsymbol{z}^t\mid N,\boldsymbol{\pi}^t\right)\propto \prod_{k=1}^{K}{\pi_{k}^t}^{M_k^{t}}$, where $M_k^{t}$ is the number of $\left\{z_{i}^t\right\}_{i=1}^{M^t}$ assigned to cluster $k$. Therefore, the conditional posterior distribution is a Dirichlet distribution:
    \begin{equation}\label{sample_pi}
    p\left(\boldsymbol{\pi}^t\mid \boldsymbol{z}^t,\boldsymbol{\alpha}\right)\sim \text{Dirichlet}\left(M_1^t+\alpha_1,M_2^t+\alpha_2,\cdots,M_K^t+\alpha_K\right).
    \end{equation}

    \item Sample $z_i^t$ from $p\left(z_i^t\mid \boldsymbol{\pi}_i^t,\boldsymbol{\mu},\boldsymbol{\Sigma},\boldsymbol{x}_i^t\right)$. The conditional distribution of assignment for each observation is given by $p\left(z_i^t=k\mid \boldsymbol{\pi}_i^t,\boldsymbol{\mu},\boldsymbol{\Sigma},\boldsymbol{x}_i^t\right) \propto p\left(z_i^t=k\mid \boldsymbol{\pi}^{t}\right) p\left(\boldsymbol{x}_i^t\mid \boldsymbol{\mu},\boldsymbol{\Sigma}\right)=\pi_k^t\mathcal{N}\left(\boldsymbol{x}_i^t\mid \boldsymbol{\mu}_k,\boldsymbol{\Sigma}_k\right)$. Then we can do the normalization to obtain the conditional discrete distribution over $z_i^t$ by \begin{equation}\label{sample_zi}
    p\left({z_{i}^t=k}\mid\boldsymbol{\pi}^t,\boldsymbol{\mu},\boldsymbol{\Sigma},\boldsymbol{x}_{i}^t\right)= \frac{\pi_{k}^t\mathcal{N}\left(\boldsymbol{x}_{i}^t\mid \boldsymbol{\mu}_k,\boldsymbol{\Sigma}_k\right)}{\sum_{m=1}^K\pi_{m}^t\mathcal{N}\left(\boldsymbol{x}_{i}^t\mid \boldsymbol{\mu}_m,\boldsymbol{\Sigma}_m\right)}.
    \end{equation}

    \item Sample $(\boldsymbol{\mu}_k,\boldsymbol{\Sigma}_k)$ from $p\left(\boldsymbol{\mu}, \boldsymbol{\Sigma} \mid \mathcal{X}_k, \Theta\right)$. Thanks to the conjugate prior distribution, the conditional distribution of the mean vector and the covariance matrix
$p\left(\boldsymbol{\mu}_k, \boldsymbol{\Sigma}_k \mid \mathcal{X}_k, \Theta \right)$ is a Gaussian-inverse-Wishart distribution:
\begin{equation}\label{mu123}
p\left(\boldsymbol{\mu}_k,\boldsymbol{\Sigma}_k \mid \mathcal{X}_k, \Theta \right) \sim \mathcal{N}\left(\boldsymbol{\mu}_k\mid\boldsymbol{\mu}_0^*,\frac{1}{\lambda_0^*}\boldsymbol{\Sigma}_k\right)\mathcal{W}^{-1}\left(\boldsymbol{\Sigma}_k\mid\boldsymbol{\Psi}_0^*,\nu_0^*\right),
\end{equation}
where
\begin{equation}\label{l123}
\begin{aligned}
\boldsymbol{\mu}_0^* = \frac{\lambda_0\boldsymbol{\mu}_0+M_k\bar{\boldsymbol{x}}}{\lambda_0+M_k}, \qquad \lambda_0^* = \lambda_0+M_k, \qquad \nu_0^*=\nu_0+M_k, \qquad \bar{\boldsymbol{x}}=\frac{1}{M_k}\sum_{t=1}^{T}\sum_{i=1}^{M_k^t}\boldsymbol{x}_{i}^t,\\
M_k = \sum_{t=1}^TM_k^t,\quad
\boldsymbol{\Psi}_0^* = \boldsymbol{\Psi}_0+\boldsymbol{\mathrm{S}}+\frac{\lambda_0M_k}{\lambda_0+M_k}\left(\bar{\boldsymbol{x}}-\boldsymbol{\mu_0}\right)\left(\bar{\boldsymbol{x}}-\boldsymbol{\mu_0}\right)^{\top}, \quad \boldsymbol{\mathrm{S}}=\sum_{t=1}^{T}\sum_{i=1}^{M_k^t}\left(\boldsymbol{x}_{i}^t-\bar{\boldsymbol{x}}\right)\left(\boldsymbol{x}_{i}^t-\bar{\boldsymbol{x}}\right)^{\top}.
\end{aligned}
\end{equation}
    \item Sample $\mathcal{X}$ from $p\left(\mathcal{X} \mid \boldsymbol{\mu}, \boldsymbol{\Sigma},\boldsymbol{z}, \mathcal{R}, \mathcal{G}  \right)$. At this step, we no longer have a simple analytical formulation to sample $\mathcal{X}$ due to the linear constraints in Eq.~\eqref{eq:linear}. Here, the conditional distribution can be factorized as
\begin{equation}\label{eq:factor}
    p\left(\mathcal{X} \mid \boldsymbol{\mu}, \boldsymbol{\Sigma},\boldsymbol{z}, \mathcal{R}, \mathcal{G}  \right) = \prod_{t=1}^{T}\prod_{i=1}^{M^t}{p\left(\boldsymbol{x}_{i}^t \mid \boldsymbol{\mu}_{z_{i}^t}, \boldsymbol{\Sigma}_{z_{i}^t}, \boldsymbol{r}_{i}^t, {\mathbf{G}}_{i}^t \right)}.
\end{equation}
Consequently, we can draw sample of the bus-pair vector $\boldsymbol{x}_{i}^t$ independently. The conditional distribution of $\boldsymbol{x}_{i}^t$ in Eq.~\eqref{eq:factor} can be regarded as a multivariate Gaussian distribution truncated on the intersection with a hyperplane, i.e.,
\begin{equation}
    \boldsymbol{x}_{i}^t\,|\, z_{i}^t=k \sim \mathcal{N}_{\mathcal{S}_{i}^t}\left(\boldsymbol{\mu}_{k},\boldsymbol{\Sigma}_{k}\right), \quad \mathcal{S}_{i}^t=\left\{\boldsymbol{x}_{i}^t\mid {\mathbf{G}}_{i}^t \boldsymbol{x}_{i}^t = \boldsymbol{r}_{i}^t \right\}.
\end{equation}
The probability density function of the hyperplane-truncated multivariate Gaussian is
\begin{equation}\label{eq:truncated_pdf}
    p(\boldsymbol{x}_{i}^t \mid \boldsymbol{\mu}_{k}, \boldsymbol{\Sigma}_{k},\boldsymbol{r}_{i}^t, \mathbf{G}_{i}^t)=\frac{1}{Z_{i}^t} \exp \left[-\frac{1}{2}(\boldsymbol{x}_{i}^t-\boldsymbol{\mu}_{z_{i}^t})^{T} \boldsymbol{\Sigma}_{z_{i}^t}^{-1}(\boldsymbol{x}_{i}^t-\boldsymbol{\mu}_{z_{i}^t})\right] \delta(\mathbf{G}_{i}^t \boldsymbol{x}_{i}^t=\boldsymbol{r}_{i}^t),
\end{equation}
where $Z_{i}^t$ is a normalizing constant; $\delta(*)$ is a function whose value is 1 if the condition $*$ holds, and 0 otherwise.
Similar to \citep{chen2022bayesian}, we adopt a fast sampling algorithm developed by \citet{cong2017fast} for this problem. The algorithm for sampling the bus-pair vector $\boldsymbol{x}_{i}^t$ is described in Algorithm~\ref{imputation}.
\end{itemize}

\begin{algorithm}
\caption{Sampling from a hyperplane-truncated multivariate Gaussian distribution \citep{cong2017fast}.}
\label{imputation}
\begin{algorithmic}[1]
\STATE Sample $\boldsymbol{u} \sim \mathcal{N}\left(\boldsymbol{\mu}_{z_{i}^t},\boldsymbol{\Sigma}_{z_{i}^t}\right)$;
\STATE Return $\boldsymbol{x}_{i}^t = \boldsymbol{u} + {\boldsymbol{\Sigma}_{z_{i}^t}}{{\mathbf{G}}_{i}^t}^{\top}\left({\mathbf{G}}_{i}^t{\boldsymbol{\Sigma}_{z_{i}^t}}{{\mathbf{G}}_{i}^t}\right)^{-1}\left(\boldsymbol{r}_{i}^t-{\mathbf{G}}_{i}^t\boldsymbol{u}\right)$, which can be more efficiently and accurately calculated by\\
\begin{itemize}
    \item Solve $\boldsymbol{\beta}$ such that $\left({\mathbf{G}}_{i}^t{\boldsymbol{\Sigma}_{z_{i}^t}}{\mathbf{G}_{i}^t}^{\top}\right)\boldsymbol{\beta}=\boldsymbol{r}_{i}^t-{\mathbf{G}}_{i}^t\boldsymbol{u}$;
    \item Return $\boldsymbol{x}_{i}^t=\boldsymbol{u}+{\boldsymbol{\Sigma}_{z_{i}^t}}{{\mathbf{G}}_{i}^t}^{\top}\boldsymbol{\beta}$.
\end{itemize}
\end{algorithmic}
\end{algorithm}

Finally, we summarize the Gibbs sampling procedure for estimating the parameters in Algorithm~\ref{alg:gibbs}. We drop the first $d_1$ iterations as burn-in and then store samples of parameters $\boldsymbol{\pi}^t$, $\boldsymbol{\mu}_k$, $\boldsymbol{\Sigma}_k$ from the last $d_2$ iterations. In particular, these stored samples mixing coefficients $\left\{{\boldsymbol{\pi}^{t}}^{(\rho)}\right\}_{\rho=1}^{d_2}$, mean vectors $\left\{\boldsymbol{\mu}_k^{(\rho)}\right\}_{\rho=1}^{d_2}$, and covariance matrices $\left\{\boldsymbol{\Sigma}_k^{(\rho)}\right\}_{\rho=1}^{d_2}$ are critical ingredients for deriving the posterior distribution of the parameters and performing probabilistic forecasting of bus travel time. For hyperparameters, we set $\boldsymbol{\mu}_0 = \boldsymbol{0}_{3n}$, $\lambda_0=10$,  $\boldsymbol{\Phi}_0 = \boldsymbol{I}_{3n}$, $\nu_0 = 3n+2$, $\boldsymbol{\alpha} = \boldsymbol{0.2}_{3n}$, where $n$ is the number of bus links.
\begin{algorithm}[H]
\renewcommand{\algorithmicrequire}{\textbf{Input:}}
\renewcommand{\algorithmicensure}{\textbf{Output:}}
\caption{Gibbs sampling for parameter estimation.}\label{alg:gibbs}
\begin{algorithmic}[1]
\REQUIRE Recording vectors $\mathcal{R}$, alignment matrices $\mathcal{G}$, hyperparameters $\Theta$ and $\boldsymbol{\alpha}$, iterations $d_1$, $d_2$.
\ENSURE Samples of mixture weights $\left\{{\boldsymbol{\pi}^t}^{(\rho)}\right\}_{\rho=1}^{d_2}$, samples of mean vectors $\left\{\boldsymbol{\mu}_k^{(\rho)}\right\}_{\rho=1}^{d_2}$, and samples of covariance matrices $\left\{\boldsymbol{\Sigma}^{(\rho)}_k\right\}_{\rho=1}^{d_2}$.
\FOR{$\mathrm{iter}=1$ to $d_1+d_2$}
\FOR{$t=1$ to $T$}
\STATE Draw $\boldsymbol{\pi}^t$ according to Eq.~\eqref{eq:pi prior}.
\IF{$\mathrm{iter}>d_1$}
\STATE Collect $\boldsymbol{\pi}^t$ to the output set.
\ENDIF
\ENDFOR
\FOR{$k=1$ to $K$}
\STATE Draw $\boldsymbol{\Sigma}_k$ and $\boldsymbol{\mu}_k$ according to Eq.~\eqref{eq:cov prior} and Eq.~\eqref{eq:mean prior}.
\IF{$\mathrm{iter}>d_1$}
    \STATE Collect $\boldsymbol{\mu}_k$ and $\boldsymbol{\Sigma}_k$ to the output sets.
\ENDIF
\ENDFOR
\FOR{$t=1$ to $T$}
\FOR{$s=1$ to $M_t$}
    \STATE Calculate p($z_{s}^t$) according to Eq.~\eqref{sample_zi}.
    \STATE Draw $z_{s}^t$ according to Eq.~\eqref{eq:zpdf}.
    \STATE Draw $\boldsymbol{x}_{s}^t$ by Algorithm~\ref{imputation}.
\ENDFOR
\ENDFOR
\FOR{$k=1$ to $K$}
\STATE Update the parameters $\Theta=\{\boldsymbol{\mu}_0,\lambda_0, \boldsymbol{\Psi}_0, \nu_0\}$ by Eq.~\eqref{l123}.
\ENDFOR
\STATE Update the parameters $\boldsymbol{\alpha}$ by Eq.~\eqref{sample_pi}.
\ENDFOR
\RETURN{$\left\{{\boldsymbol{\pi}^t}^{(\rho)}\right\}_{\rho=1}^{d_2}$,$\left\{\boldsymbol{\mu}^{(\rho)}_k\right\}_{\rho=1}^{d_2}$, $\left\{\boldsymbol{\Sigma}^{(\rho)}_k\right\}_{\rho=1}^{d_2}$.}
\end{algorithmic}
\end{algorithm}

\subsection{Probabilistic Forecasting}\label{sec:forecast}
We divide the links of each bus into observed and upcoming links at the time of making forecasting. Observed links are passed links with known travel times, upcoming links are future links whose link travel times are to be predicted. Because of the Gaussian assumption in each mixture component, we can easily forecast the bus travel time on upcoming links by calculating the conditional distribution given observed link travel times and headways. Generally, there are upcoming links for both the following bus and the leading bus; although it is possible to make predictions conditioning on only the observed links, we adopt an autoregressive approach that also uses the forecasting of the leading bus's upcoming links to forecast the bus travel time of the following bus, because this autoregressive approach uses the information of the leading bus's leading bus to reinforce the forecasting.
For the first bus (without a leading bus) of a day, we only use observed links to forecast the upcoming links (like the method proposed by \citet{chen2022bayesian}).

\begin{figure}[htbp]
    \centering
    \subfigure{\includegraphics[width = 0.85\textwidth]{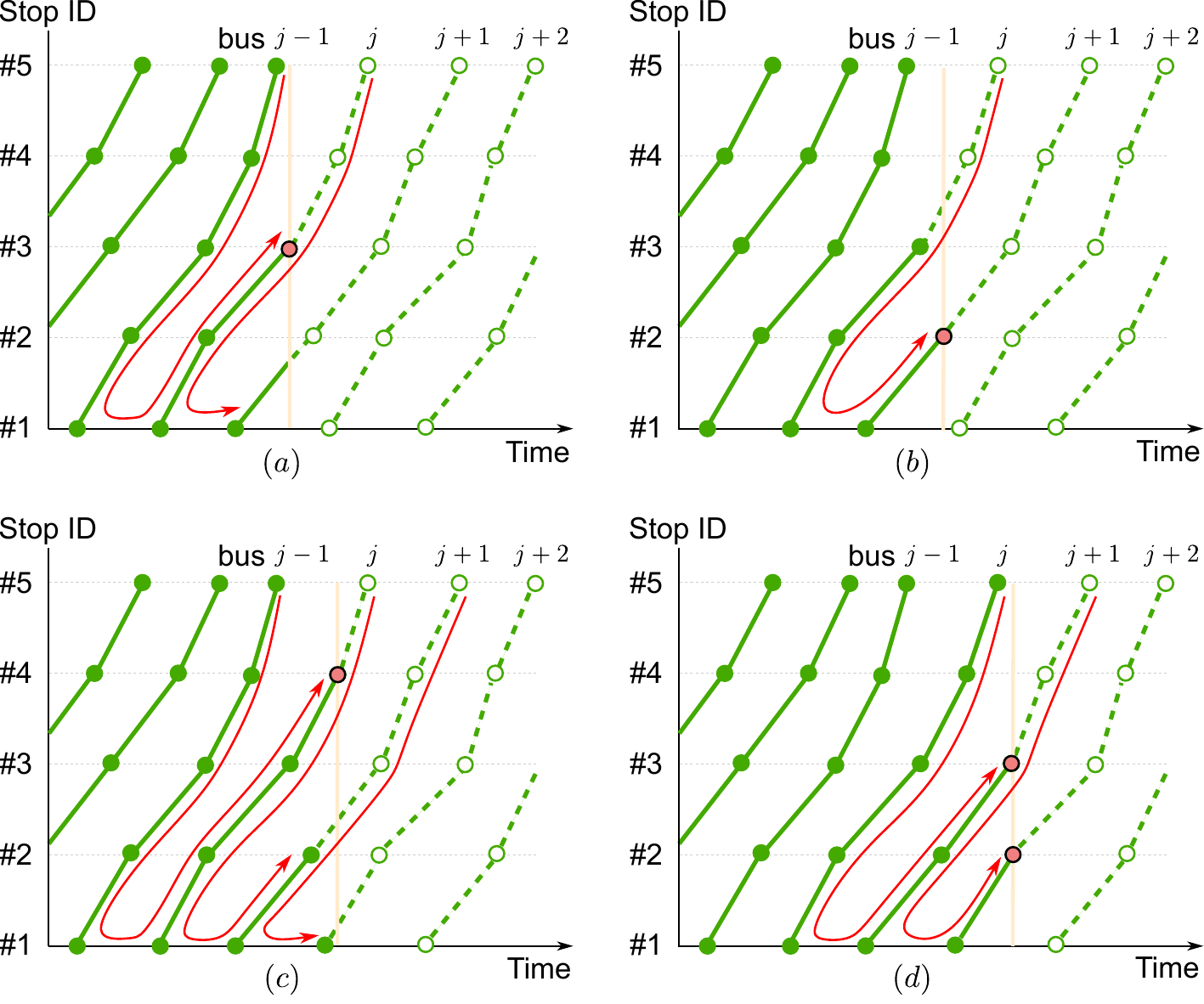}}
    \caption{Illustration of travel time forecasting. Solid dots: observed arrival time at bus stations. The hollow dot: unknown arrival time at the bus station. Red solid dots: new observed arrival time at bus stations. Each sub-figure is a new round of forecasting triggered by a new observation of bus arrival. For each adjacent bus pair, we forecast the following bus's upcoming link travel time (dash green lines) based on observed and previously predicted link travel times (green lines covered by a red arrow area).} \label{fig:forecast_example}
\end{figure}

We use the Fig.~\ref{fig:forecast_example} to illustrate the forecasting process.
In Fig.~\ref{fig:forecast_example}(a), bus $j-1$ has just finished the run; bus $j$ has passed the first two links and arrived at the stop $\#3$; bus $j+1$ has departed from the origin stop but does not arrive at the stop $\#2$. We would like to use the observed links/headways to make forecasting for upcoming links of the bus $j$ and the bus $j+1$. For the bus $j$, we could make forecasting by using its upstream links (the first two links), all the observed link travel times of the leading bus $j-1$, and corresponding observed headways (the first three headways). In terms of the bus $j+1$, we could use the observed upstream link travel times of the bus $j$, the forecasts of the upcoming link travel times of the bus $j$, and the observed headway. As the time passing, related buses could update the forecasting once having new observed links. Fig.~\ref{fig:forecast_example}(b) shows that the first new observed link comes up after the time point in Fig.~\ref{fig:forecast_example}(a). We can see that at this time point, only bus $j+1$ gets a new observed link; thus we could update the forecasting of bus $j+1$. At the next time point shown in Fig.~\ref{fig:forecast_example}(c), bus $j$ gets a new observed link; thus, bus $j$ need to update the forecasting; although bus $j+1$ does not have new observed links, the updated information of its leading bus (i.e., bus $j$) could reinforce its forecasting. In addition, we can see a new bus $j+2$ begins to run on the route; with the observed headway, we could make forecasting for bus $j+2$. Next, in Fig.~\ref{fig:forecast_example}, bus $j$ have finished the run; bus $j+1$ and $j+2$ get a new observed link, respectively. In this case, we could make forecasting for bus $j+1$ and $j+2$ in the same way.

For the $j$-th bus pair with a following bus started at the $t$-th period in a day, assume at current time point $t^*$ we can observe the first $p$ links of the leading bus, the first $q$ links of the following bus, and the corresponding headways, then we aim to forecast $\hat{\boldsymbol{x}}^{t,t^*}_{j}$ for the upcoming links of the following bus.   Apart from the observed information, we also combine the forecasts $\hat{\boldsymbol{x}}^{t_l,t^*}_{j-1}$ of the leading bus to construct the recording vector $\boldsymbol{r}_{j}^{t,t*}$ and alignment matrix $\mathbf{G}_{j}^{t,t^*}$. The posterior predictive distribution of the augmented random variable  $\boldsymbol{x}_{j}^{t,t^*}$ is
\begin{equation}\label{eq:postpre}
\begin{aligned}
    p(\boldsymbol{x}_{j}^{t,t^*} \mid \boldsymbol{r}_{j}^{t,t^*}, \mathbf{G}_{j}^{t,t^*}) = \iiiint p(\boldsymbol{x}_{j}^{t,t^*} \mid \boldsymbol{\mu}, \boldsymbol{\Sigma}, \boldsymbol{z}^t,\boldsymbol{r}_{j}^{t,t^*}, \mathbf{G}_{j}^{t,t^*})p(\boldsymbol{\mu},\boldsymbol{\Sigma}\mid \Theta)p(\boldsymbol{z}^t\mid \boldsymbol{\pi}^t)p(\boldsymbol{\pi}^t\mid\boldsymbol{\alpha})\,d\boldsymbol{\mu}\,d\boldsymbol{\Sigma}\,d\boldsymbol{z}^t\,d\boldsymbol{\pi}^t.
\end{aligned}
\end{equation}
From the joint distribution, we can easily obtain the posterior predictive distribution $p\left(\hat{\boldsymbol{x}}_{j}^{t,t^*}\right)$ over the upcoming $(n-q)$ links by sampling methods. Note that we have gathered samples of parameters in the procedure of model inference; thus, we could directly use the stored samples to make probabilistic forecasting, which could avoid the time cost of parameters sampling. We summarize the procedure of probabilistic forecasting in Algorithm~\ref{alg:gibbs4cast}.

\begin{algorithm}
\renewcommand{\algorithmicrequire}{\textbf{Input:}}
\renewcommand{\algorithmicensure}{\textbf{Output:}}
\caption{Gibbs sampling for probabilistic forecasting.}\label{alg:gibbs4cast}
\begin{algorithmic}[1]
\REQUIRE Observed vector $\boldsymbol{r}_{j}^{t,t^*}$, alignment matrices $\mathbf{G}_{j}^{t,t^*}$, samples of mixture weights $\left\{{\boldsymbol{\pi}^t}^{(\rho)}\right\}_{\rho=1}^{d_2}$,  samples of covariance matrices $\left\{\boldsymbol{\Sigma}^{(\rho)}_k\right\}_{\rho=1}^{d_2}$, samples of mean vectors $\left\{\boldsymbol{\mu}^{(\rho)}_k\right\}_{\rho=1}^{d_2}$.
\ENSURE A set of samples for the forecast $\hat{\boldsymbol{x}}_{j}^{t,t^*}$.
\FOR{$\rho=1$ to $d_2$}
\STATE Compute $p(z_{\rho}^t)$ according to Eq.~\eqref{sample_zi}.
\STATE Draw $z_{\rho}^t$ according to Eq.~\eqref{eq:zpdf}.
\STATE Draw $\boldsymbol{x}_{j}^{t,t^*}$ by Algorithm~\ref{imputation}.
\STATE Collect $\hat{\boldsymbol{x}}_{j}^{t,t^*}$ to the output set.
\ENDFOR
\RETURN{$\left\{{\hat{\boldsymbol{x}}_{j}^{{t,t^*}^{(\rho)}}}\right\}^{d_2}_{\rho=1}$.}
\STATE Get the posterior predictive distributions from samples $\left\{{\hat{\boldsymbol{x}}_{j}^{{t,t^*}^{(\rho)}}}\right\}^{d_2}_{\rho=1}$.
\end{algorithmic}
\end{algorithm}

\section{Experiments}
\label{casestudy}
\subsection{Data and Experimental Settings}
\begin{figure*}[!ht]
\centering
\subfigure{
    \centering
    \includegraphics[width=0.78\textwidth]{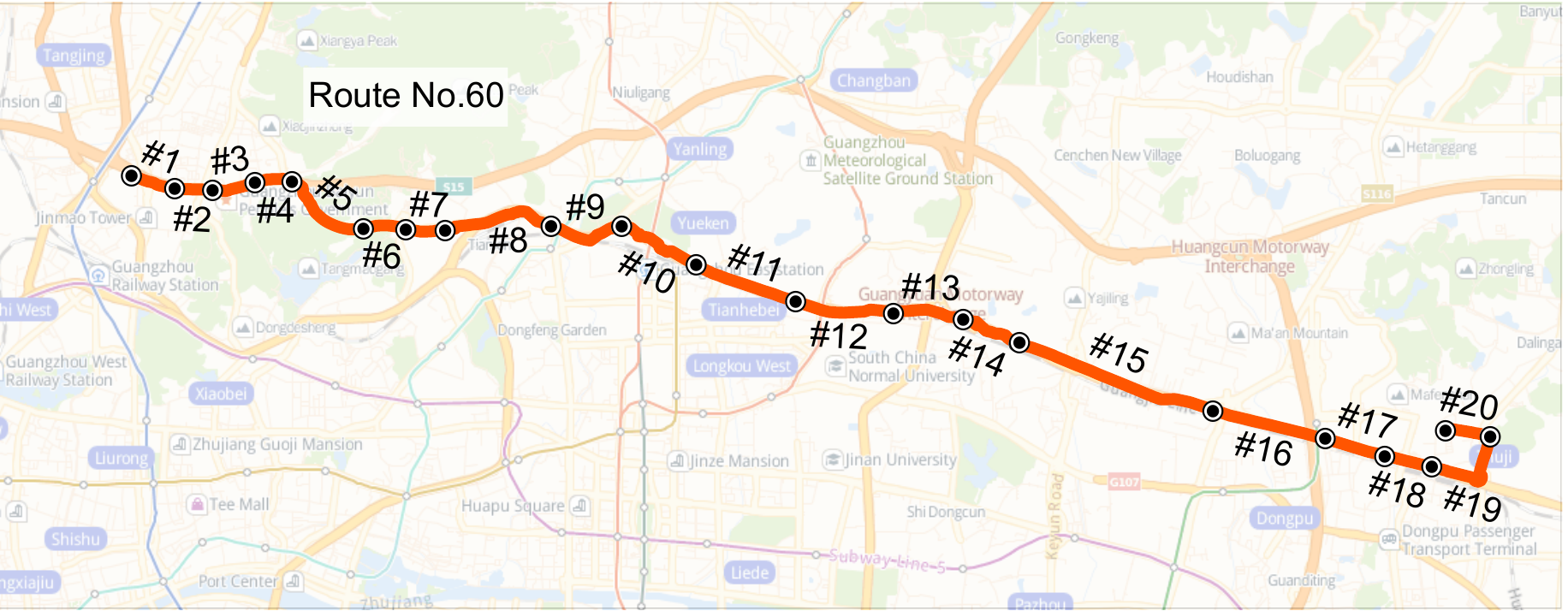}
}
\caption{Bus route No. 60 in Guangzhou bus network.}
\label{Bus_network}
\end{figure*}

\begin{figure*}[!ht]
\centering
\subfigure{
    \centering
    \includegraphics[width=0.79\textwidth]{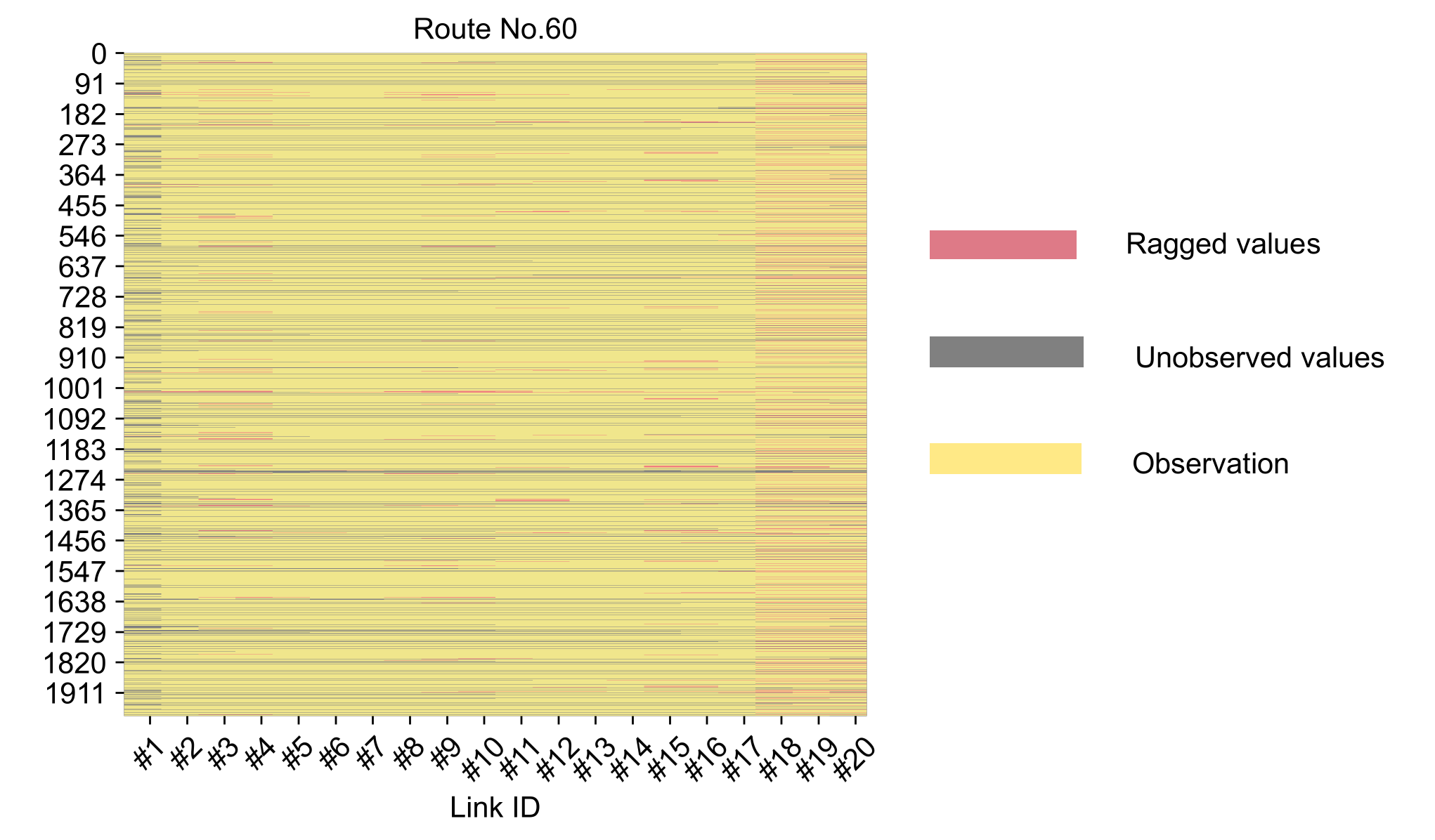}
}
\caption{Data overview.}
\label{data_overview}
\end{figure*}

In our experiments, we evaluate the proposed probabilistic forecasting model on the real-world data. The data used in this paper are the bus in-out-stop record data collected in Guangzhou, China, during the weekdays from December 1st, 2016 to December 31st, 2016. The information of the data is outlined in Table~\ref{desciption_data}. These data were collected by the bus in-out-stop record system, i.e., the automatic bus announcing system. When a bus enters or exits a bus stop, the system reports the arrival or departure information and records the timestamp accordingly. Thus we can easily obtain the link travel times/headways from the data. We take bus route No. 60 as a case and aim to make probabilistic forecasting for travel time of this route. This bus route has 21 stops and 20 links, which are displayed in Fig.~\ref{Bus_network}, and they are in the urban areas of Guangzhou. The overview of data is shown in Fig.~\ref{data_overview}. We can see that the bus route has many missing and ragged values. Moreover, Fig.~\ref{Empdis} shows the empirical distributions of link travel times of route No. 60. We can see that many link travel times exhibit positively skewed and unimodal distributions while some links such as link $\#11$ and link $\#12$ have bimodal distributions. In this paper, we thus use the Gaussian mixture model to approximate the skewed and multimodal distributions.

\begin{table}[!ht]
\caption{Description of Data.}
\label{desciption_data}
\centering
\footnotesize
\begin{tabular}{l|l|l}
\toprule
Variable & Description & Example\\
\midrule
ID & Identity for bus data record & 1612020547101390 \\
OBUID & Identity for bus & 911721 \\
TRIP\_ID & Identity for bus trip & 1612012250030880 \\
ROUTE\_ID & Identity for bus route & 201 \\
ROUTE\_NAME & Bus route name & No. 24 \\
ROUTESUB\_ID & Identity of bus route direction & 502669 \\
ROUTE\_STA\_ID & Identity of bus stop & 84279 \\
STOP\_NAME & Bus stop name & Dunhe Stop \\
AD\_FLAG & Bus state: arrival (1) or departure (0) & 1\\
AD\_TIME & The time bus reported arriving at/leaving a bus stop & 20161202, 05:47:08\\
\bottomrule
\end{tabular}
\end{table}

As the measurements have different units, we could do the standardization (z-score normalization) so that they are centered around 0 with a standard deviation of 1. By doing so, we can learn the residual covariance (zero-centered) instead of the result of the large mean. For example, the data $\ell_{i,j}$ (the $j$-th link travel time of the $i$-th bus) can be rescaled with
\begin{equation}
    \Tilde{\ell}_{i,j} =  \frac{\ell_{i,j}-\mu_{\ell_j}}{\sigma_{\ell_j}},
\end{equation}
where $\mu_{\ell_j}$ is the mean of travel time at the $j$-th link; $\sigma_{\ell_j}$ is the standard deviation of travel time at the $j$-th link. Note that the constraints in Eq.~\eqref{eq:headway_constraint} no longer hold after the standardization (e.g., $\Tilde{\ell}_{i,2} + \Tilde{\ell}_{i,3} \neq r_{i,2}, \Tilde{h}_{i,3}-\Tilde{h}_{i,2}+\Tilde{\ell}_{i,2}-\Tilde{\ell}_{i-1,2} \neq 0$). Therefore, we need to transform the constraints to ensure the equality; luckily, this can be achieved by a simple linear transform.
As we estimate the model parameters by using standardizing data, the forecasts could be recovered by the transformation $\hat{\ell}_{i,j}=\hat{\Tilde{\ell}}_{i,j}\sigma_{\ell_j}+\mu_{\ell_j}$, where $\hat{\Tilde{\ell}}_{i,j}$ is the forecast of standardizing data.
We use the first sixteen weekdays (December 1st to December 23rd) for model estimation, and the following five weekdays (December 26th to December 30th) to test the forecasting model. In this experiment, we make probabilistic forecasting for the link/trip travel time to evaluate the proposed model. The numbers of MCMC iterations are $d_1=9000$ (burn-in iterations) and $d_2=1000$ (sampling iterations), respectively.

\subsection{Performance Metrics}
Non-probabilistic forecasting usually applies the root mean squared error (RMSE) and the mean absolute percentage error (MAPE) to evaluate the performance. However, these two performance metrics in probabilistic forecasting can only evaluate the expectation of forecasting instead of the distribution. Therefore, we use the logarithmic score (LogS) and the continuous rank probability score (CRPS) to evaluate the probabilistic forecasting.

\begin{itemize}
    \item The RMSE and MAPE are defined as:
    \begin{equation}
    \begin{aligned}
         \text{RMSE}&=\sqrt{\frac{1}{n}\sum_{i=1}^{n}(y_i-\hat{y}_{i})^2}, \\
         \text{MAPE}&=\frac{1}{n}\sum_{i=1}^{n}\left|\frac{y_i-\hat{y}_i}{y_i}\right|,
    \end{aligned}
    \end{equation}
    where $y_i,\hat{y}_{i},i=1,\ldots,n$ are the true values and forecasts, respectively.
    \item The logarithmic score refers to likelihood and is formally defined as
    \begin{equation}
        \text{LogS}(f_{X},y) = -\log f_{X}(y),
    \end{equation}
    where $f_{X}$ is the multivariate forecasting probability density function (PDF), and $y$ is the observation. LogS is equivalent to the log-likelihood of the forecasting probability distribution and it captures all possible information about the observations related the model. Here, we use the average LogS of all observations to evaluate the model.
    \item The continuous rank probability score is often used as a quantitative measure of probabilistic forecasting; it is defined as the quadratic measure of discrepancy between the forecasting cumulative distribution function (CDF), noted $F_{X}$, and $\mathbb{I}(x\geq y)$, the empirical CDF of the observation $y$
    \begin{equation}
        \text{CRPS}(F_{X},y)=\int_{-\infty}^{\infty}\left(F_{X}(x)-\mathbb{I}(x\geq y)\right)^2dx,
    \end{equation}
    where $\mathbb{I}(\cdot)$ is the indicator function. We use the average CRPS of all observations as one metric.
\end{itemize}

\subsection{Models in Comparison}
We compare the following models to demonstrate the effect of the leading bus and headway on probabilistic forecasting for bus travel time.
\begin{itemize}
    \item Model A: all buses are independent; the model uses observed links to make conditional probabilistic forecasting.
    \item Model B: adjacent buses are dependent; the model uses the observed links of the bus pairs to make conditional forecasting without headways.
    \item Model C: adjacent buses are dependent; the model uses both the link travel times and headways of the bus pair to make conditional forecasting.
\end{itemize}

\subsection{Forecasting Results}
We apply  Algorithm~\ref{alg:gibbs} to estimate model parameters and Algorithm~\ref{alg:gibbs4cast} to make probabilistic forecasting for bus travel time. The number of clusters in the models is set as 1, 2, and 5, respectively. To evaluate the travel time forecasting with different numbers of observed links, we especially test the route with 5, 10, and 15 links, respectively. Table~\ref{tab: link} shows the forecasting performance of different models for link travel time and Table~\ref{tab: trip} shows the forecasting performance for trip travel time. Of these results, the forecasting performance of each model can be improved significantly with the increase in the number of observed links. By comparing these three models, it is not hard to see that the proposed model (Model C) outperforms both Model A and Model B, demonstrating the importance of the information on the leading bus and headways. Therefore, we can conclude that the travel time and headways of the bus pair can reinforce the probabilistic forecasting for bus travel time. Observing the model with different numbers of clusters in the Gaussian mixture model empirically demonstrates the importance of mixture for characterizing bus travel time.

\begin{table}[!htp]\centering
\caption{Performance of different models for link travel time conditional forecasting.}\label{tab: link}
\scriptsize
\resizebox{\textwidth}{!}{
\begin{tabular}{l|c|c|c|c|c|c|c|c|c|c|c|c|c}\toprule
\multicolumn{2}{c}{\multirow{3}{*}{}} &\multicolumn{12}{c}{Observed links} \\
\cmidrule{3-14}
\multicolumn{2}{c}{} &\multicolumn{4}{c|}{5 links} &\multicolumn{4}{c|}{10 links} &\multicolumn{4}{c}{15 links} \\
\cmidrule{3-14}
\multicolumn{2}{c}{} &RMSE &MAPE &CRPS &LogS &RMSE &MAPE &CRPS &LogS &RMSE &MAPE &CRPS &LogS \\
\midrule
\multirow{3}{*}{Model A} &K = 1 &33.9 &0.1439 &15.48 &-4.495 &31.8 &0.1274 &14.54 &-4.413 &27.9 &0.1151 &13.03 &-4.367  \\
&K = 2 &33.8 &0.1436 &14.90 &-4.553 &32.1 &0.1275 &14.13 &-4.698 &27.9 &0.1175 &12.55 &-4.418 \\
&K = 5 &34.1 &0.1430 &14.51 &-4.456 &32.6 &0.1252 &13.53 &-4.855 &29.6 &0.1200 &11.79 &-4.288 \\
\midrule
\multirow{3}{*}{Model B} &K = 1 &33.5 &0.1369 &15.02 &-4.451 &29.7 &0.1142 &13.26 &-4.342 &30.3 &0.1179 &13.32 &-4.344 \\
&K = 2 &33.7 &0.1442 &14.86 &-4.434 &29.3 &0.1171 &12.89 &-4.303 &31.1 &0.1233 &13.07 &-4.297 \\
&K = 5 &34.5 &0.1387 &14.51 &-4.411 &29.7 &0.1148 &12.34 &-4.261 &31.9 &0.1245 &12.12 &-4.220 \\
\midrule
\multirow{3}{*}{Model C} &K = 1 &33.0 &0.1341 &14.49 &-4.422 &29.3 &0.1139 &12.62 &-4.306 &31.9 &0.1187 &12.78 &-4.273 \\
&K = 2 &\textbf{29.7} &\textbf{0.1252} &\textbf{13.11} &\textbf{-4.334} &\textbf{22.0} & 0.0989 & 10.26 & \textbf{-4.164} & \textbf{17.0} & 0.0918 & \textbf{7.93} &\textbf{-3.970} \\
&K = 5 & 30.3 & 0.1253 &13.19 &-4.341 &22.1 &\textbf{0.0986} & \textbf{10.22} &-4.171 &17.1 & \textbf{0.0874} &7.97 &-3.990 \\
\bottomrule
\multicolumn{14}{l}{{Best results are highlighted in bold fonts.}}
\end{tabular}}
\end{table}

\begin{table}[!htp]\centering
\caption{Performance of different models for trip travel time conditional forecasting.}\label{tab: trip}
\scriptsize
\resizebox{\textwidth}{!}{
\begin{tabular}{l|c|c|c|c|c|c|c|c|c|c|c|c|c}\toprule
\multicolumn{2}{c}{\multirow{3}{*}{}} &\multicolumn{12}{c}{Observed links} \\
\cmidrule{3-14}
\multicolumn{2}{c}{} &\multicolumn{4}{c|}{5 links} &\multicolumn{4}{c|}{10 links} &\multicolumn{4}{c}{15 links} \\
\cmidrule{3-14}
\multicolumn{2}{c}{} &RMSE &MAPE &CRPS &LogS &RMSE &MAPE &CRPS &LogS &RMSE &MAPE &CRPS &LogS \\
\midrule
\multirow{3}{*}{Model A} &K = 1 &187.8 &0.0789 &110.23 &-6.921 &132.2 &0.0860 &77.42 &-6.411 &71.5 &0.0863 &43.87 &-5.845 \\
&K = 2 &188.4 &0.0790 &105.97 &-6.751 &136.3 &0.0877 &76.14 &-6.492 &70.4 &0.0855 &41.70 &-5.764 \\
&K = 5 &190.1 &0.0790 &106.96 &-6.712 &137.2 &0.0876 &73.86 &-6.320 &72.9 &0.0878 &37.24 &-5.544 \\
\midrule
\multirow{3}{*}{Model B} &K = 1 &177.4 &0.0760 &102.13 &-6.696 &119.9 &0.0762 &68.51 &-6.272 &70.9 &0.0884 &42.10 &-5.780 \\
&K = 2 &182.1 &0.0801 &105.01 &-6.709 &117.7 &0.0770 &67.24 &-6.254 &73.6 &0.0911 &41.71 &-5.720 \\
&K = 5 &185.1 &0.0786 &102.37 &-6.704 &117.0 &0.0740 &63.98 &-6.180 &74.0 &0.0908 &36.45 &-5.584 \\
\midrule
\multirow{3}{*}{Model C} &K = 1 &171.6 &0.0713 &96.51 &-6.594 &115.9 &0.0729 &64.17 &-6.151 &75.6 &0.0909 &40.41 &-5.647 \\
&K = 2 &\textbf{149.5} &\textbf{0.0686} & \textbf{83.79} & \textbf{-6.443} & 87.2 & 0.0651 & 48.46 & -5.865 & 36.0 & \textbf{0.0619} & 19.42 & -4.943 \\
&K = 5 & 151.6 &0.0694 &84.77 &-6.502 &\textbf{86.1} &\textbf{0.0641} & \textbf{47.83} & \textbf{-5.850} & \textbf{35.8} &0.0625 & \textbf{19.38} & \textbf{-4.931} \\
\bottomrule
\multicolumn{14}{l}{{Best results are highlighted in bold fonts.}}
\end{tabular}}
\end{table}

\subsection{Interpreting Mixture Components}
\begin{figure*}[!ht]
\centering
\subfigure{
    \centering
    \includegraphics[width=0.95
    \textwidth]{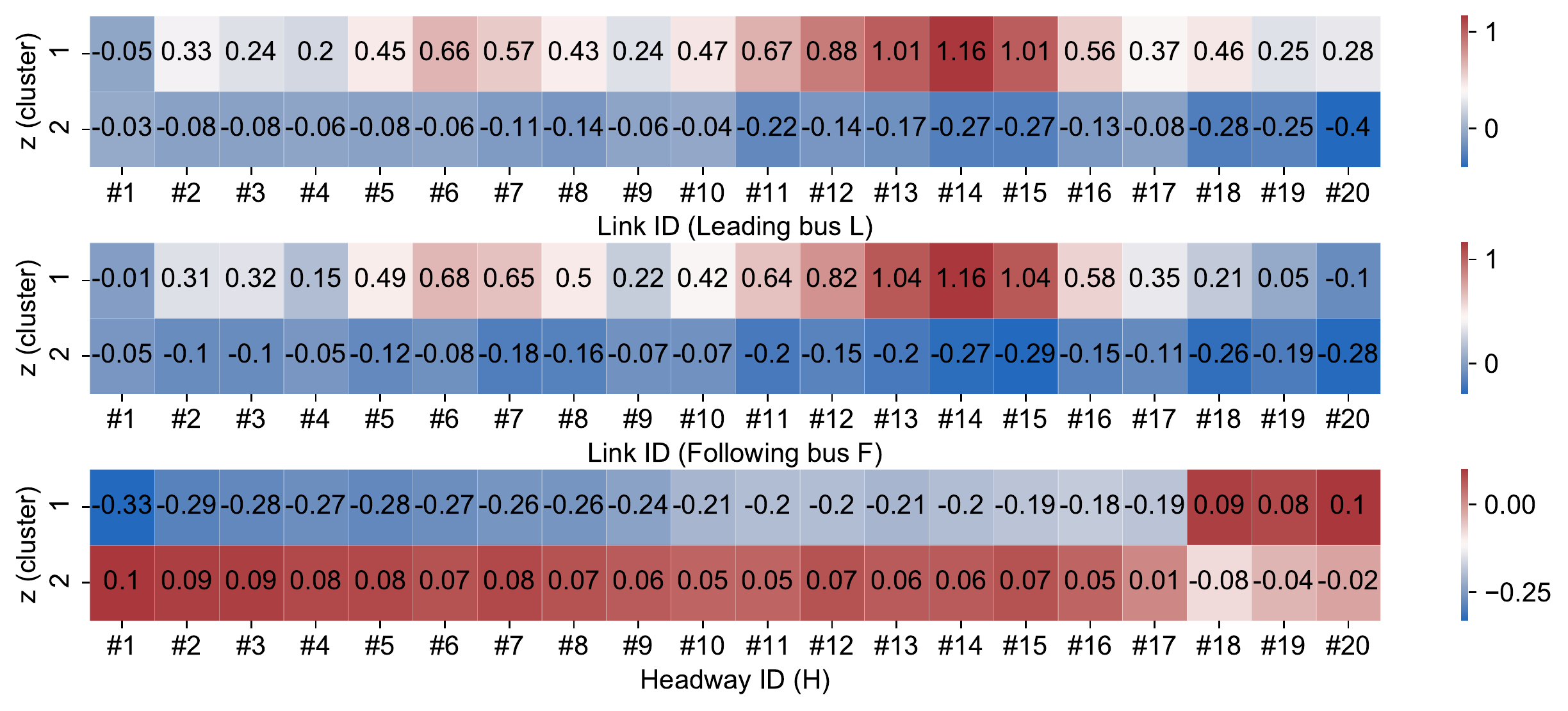}
}
\caption{Mean vectors for different clusters.}
\label{fig:mean_vector}
\end{figure*}

\begin{figure*}[!ht]
\centering
\subfigure[Trajectory samples for cluster 1.]{
    \centering
    \includegraphics[width=0.43
    \textwidth]{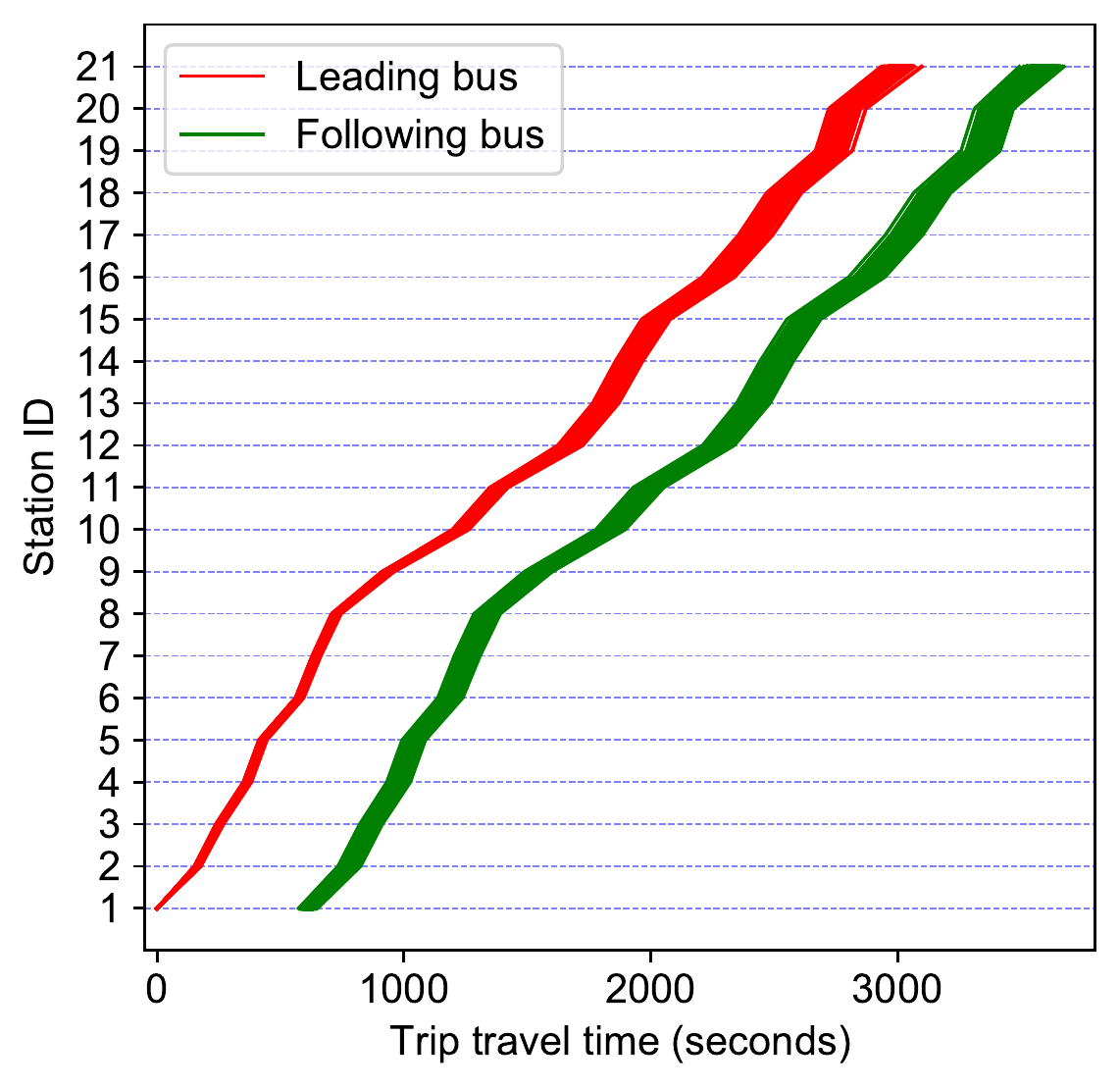}
}
\subfigure[Trajectory samples for cluster 2.]{
    \centering
    \includegraphics[width=0.43
    \textwidth]{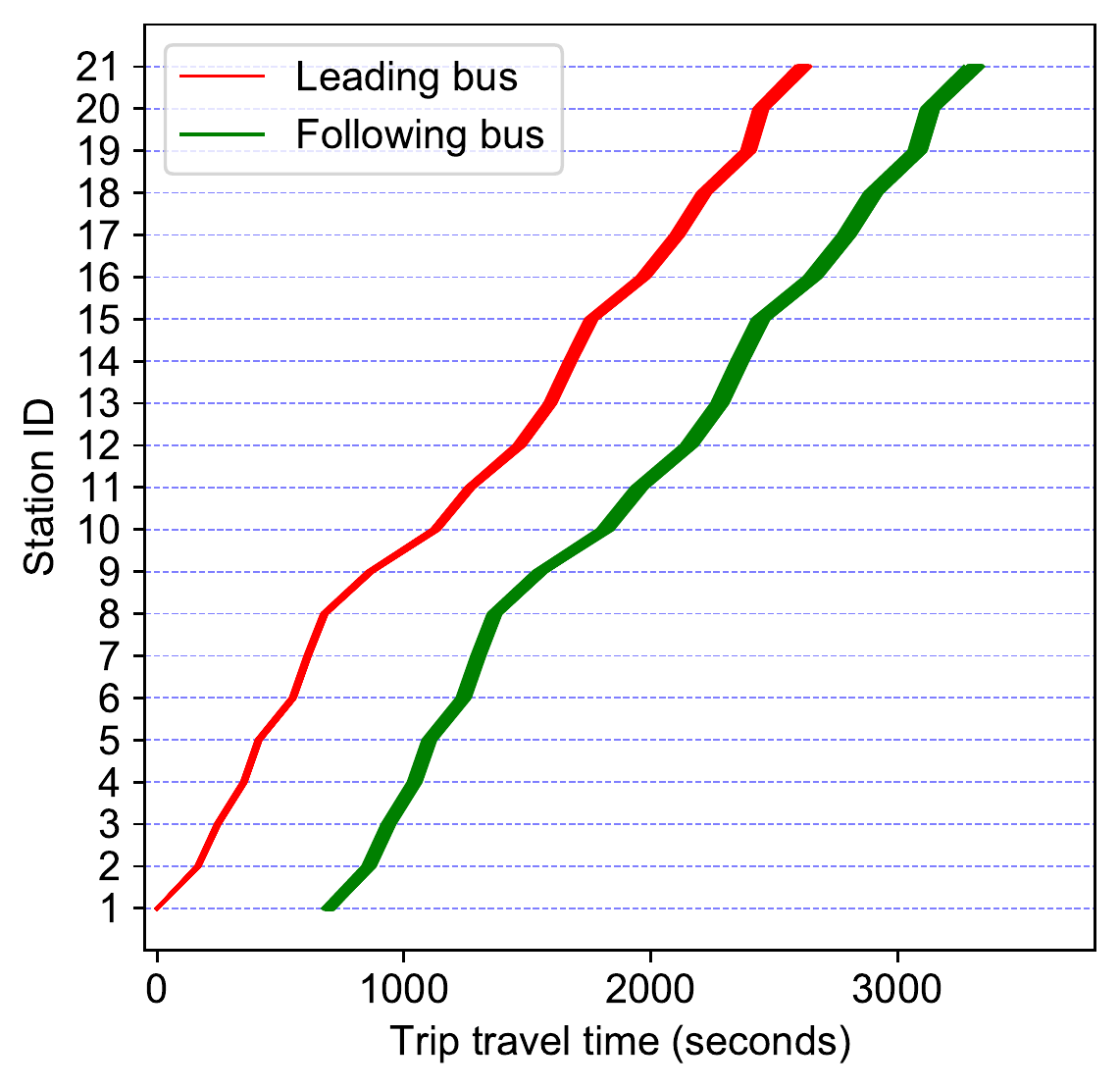}
}
\caption{Distribution of the  estimated trajectory for different clusters.}
\label{fig:cluster_trip}
\end{figure*}

After the above analysis, we use $K=2$ to show the practical implication of the probabilistic forecasting for bus travel time. Fig.~\ref{fig:mean_vector} shows the estimated mean vectors (standardization) for different clusters/patterns. We can see that the mean vectors demonstrate significant differences in some link travel times (e.g., link $\#11$, $\#12$, $\#13$, $\#14$, $\#15$) and many headways. To better find the difference of clusters in terms of link travel time/headway,  Fig.~\ref{fig:cluster_trip} visualize the trajectory plots by using the sampling link travel times and headways. By comparing the estimated trajectories of cluster 1 and cluster 2, we can find that: cluster 1 has longer link/trip travel times than cluster 2; cluster 1 has shorter headways than cluster 2; cluster 1 has larger variances for link travel times. Therefore, we can conclude that the buses of cluster 1 are more likely to run during peak hours when the heavy passenger flow and the traffic congestion often happen. On the other hand, the buses of cluster 2 might run during off-peak hours with a high probability.

\begin{figure*}[!ht]
\centering
\subfigure{
    \centering
    \includegraphics[width=0.69
    \textwidth]{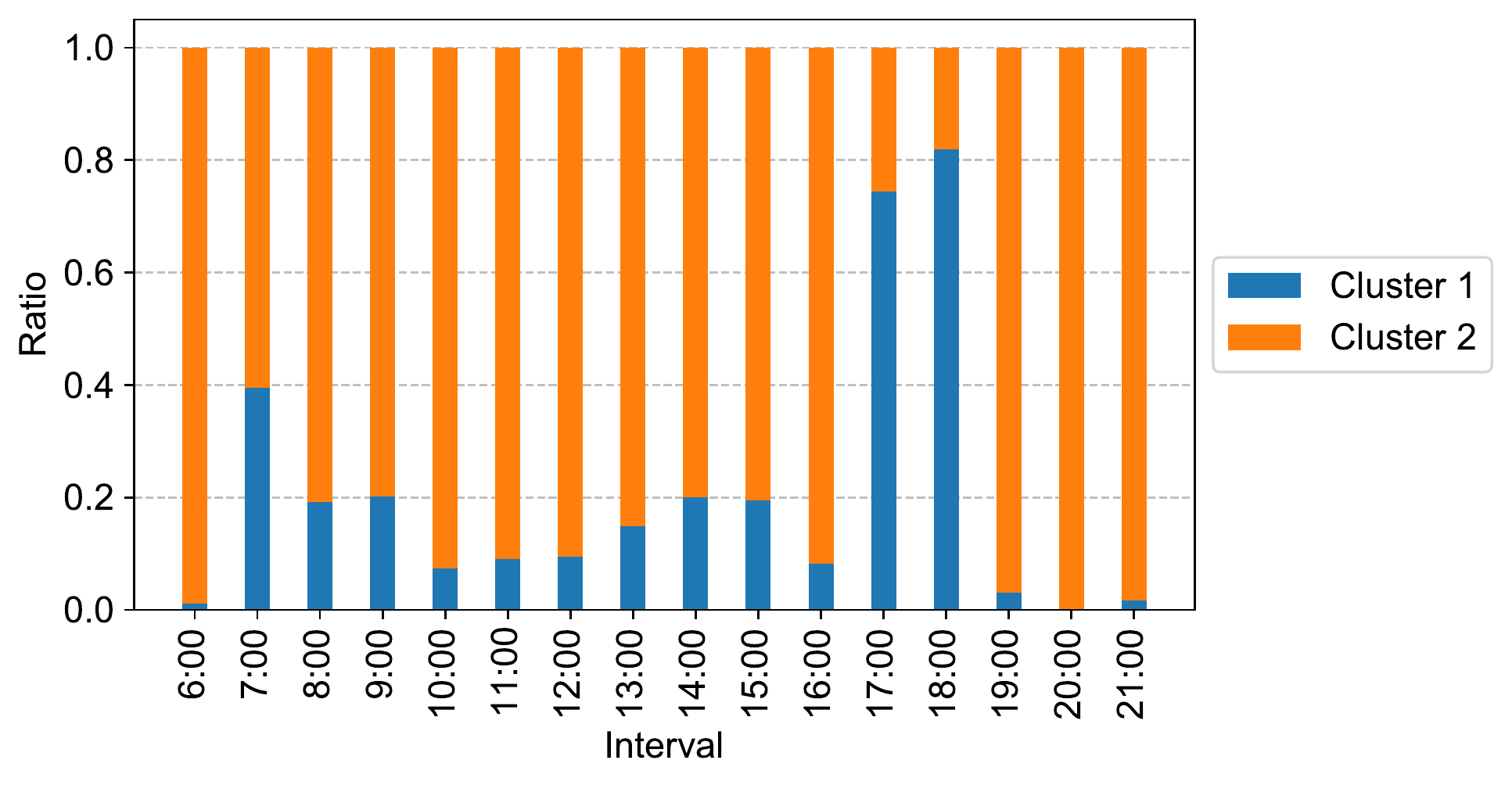}
}
\caption{Cluster distribution for different intervals.}
\label{fig:interval}
\end{figure*}

To interpret the clusters, Fig.~\ref{fig:interval} depicts the clear time-evolving patterns of each cluster. We can see that cluster 1 is dominant for afternoon peak hours while it is inferior to cluster 2 for off-peak hours and morning peak hours. Generally, morning peak and afternoon peak have similar traffic characteristics while in this case morning peak has similar characteristics to off-peak hours. This is because the studied directional bus route stretches from urban business districts to suburban areas. The better traffic condition and the small passenger flow exist in the morning because few people go to suburban areas on weekday mornings. On the contrary, traffic congestion and large passenger flow happen in the afternoon peak as more people go home from urban to suburban areas. Therefore, cluster 1 (afternoon peak) exhibits a longer travel time, shorter headway (higher frequency), and larger uncertainty than cluster 2 (morning peak, off-peak hours).

\begin{figure*}[!ht]
\centering
\subfigure[Correlation matrix for cluster 1.]{
    \centering
    \includegraphics[width=0.65
    \textwidth]{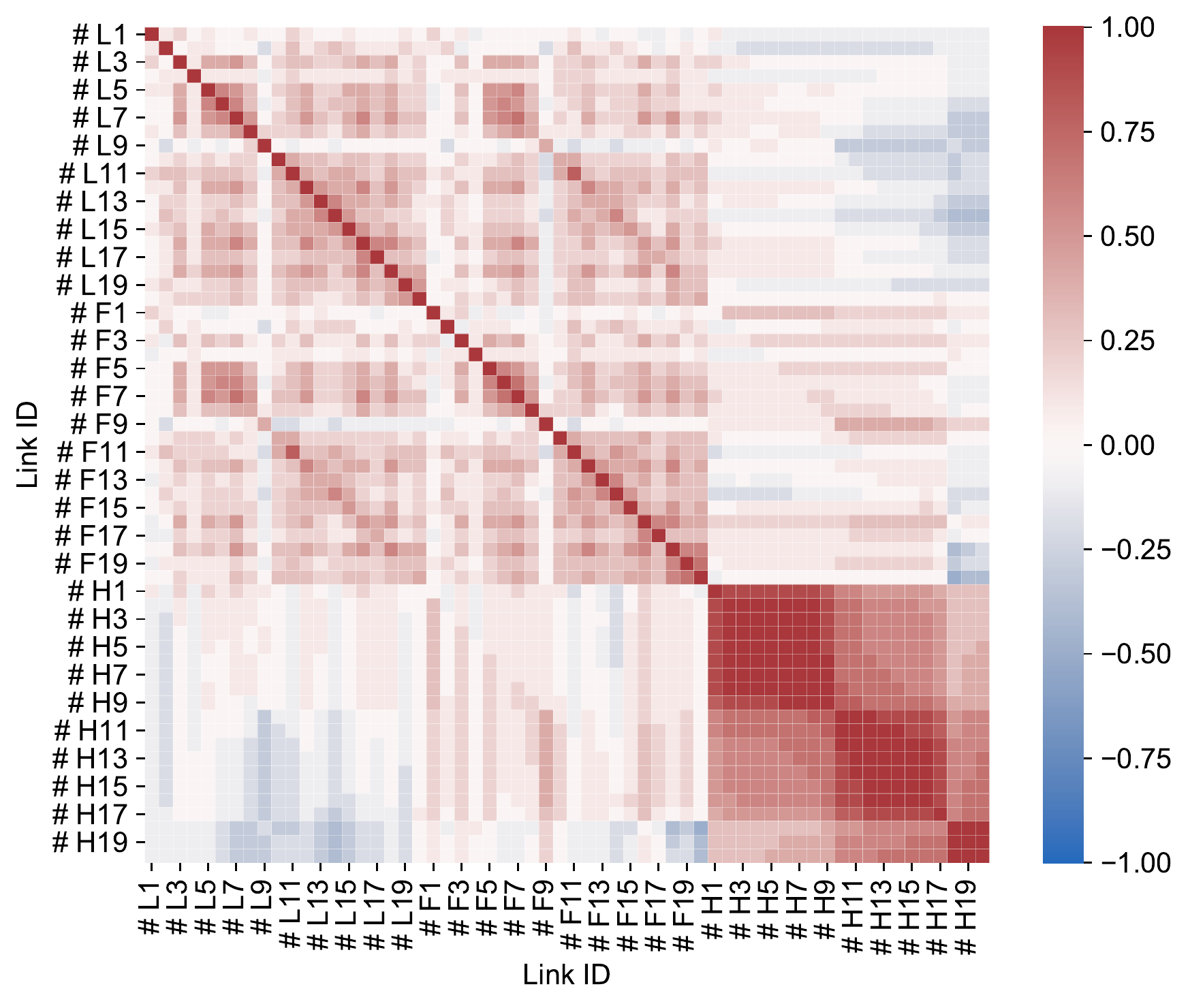}
}
\subfigure[Correlation matrix for cluster 2.]{
    \centering
    \includegraphics[width=0.65
    \textwidth]{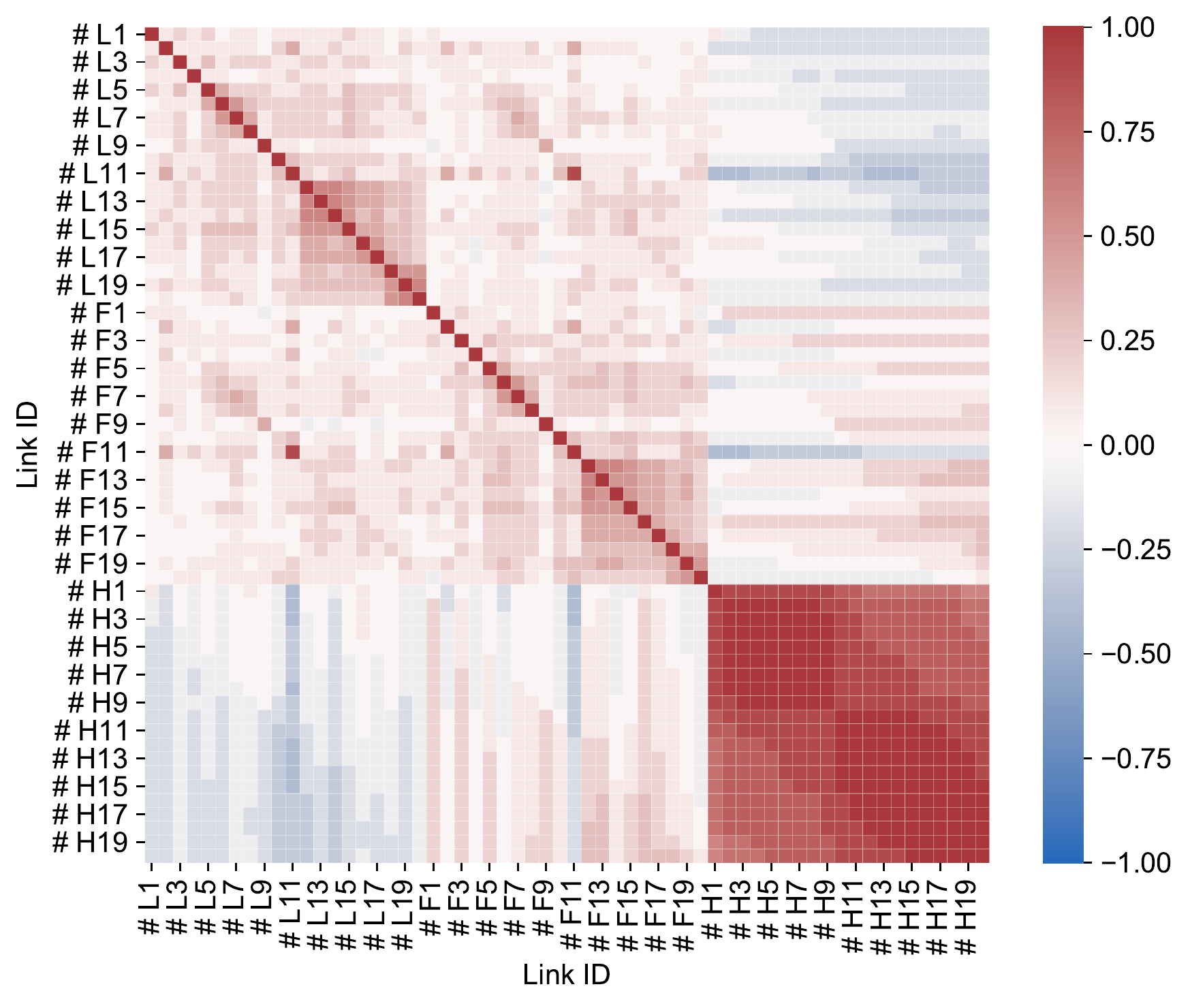}
}
\caption{Correlation matricies for different clusters.}
\label{fig:correlation}
\end{figure*}

Moreover, Fig.~\ref{fig:correlation} depicts the correlation matrices for different clusters/patterns. Both of the correlation matrices show the complex characteristics of the correlations between link travel times/headways: 1) long-range correlations, 2) negative correlations, and 3) different patterns. On the other hand, it also shows clear differences between correlation matrices of different clusters. The most significant difference is that the leading bus and the following bus for cluster 1 could be more correlated than that for cluster 2. Therefore, characterizing the bus travel time with a Gaussian mixture model is of great significance in practice.

\subsection{Distribution of the Predicted Bus Travel Time}
\begin{figure*}[!ht]
\centering
\subfigure[Probabilistic forecasting for 16:40.]{
    \centering
    \includegraphics[width=0.47
    \textwidth]{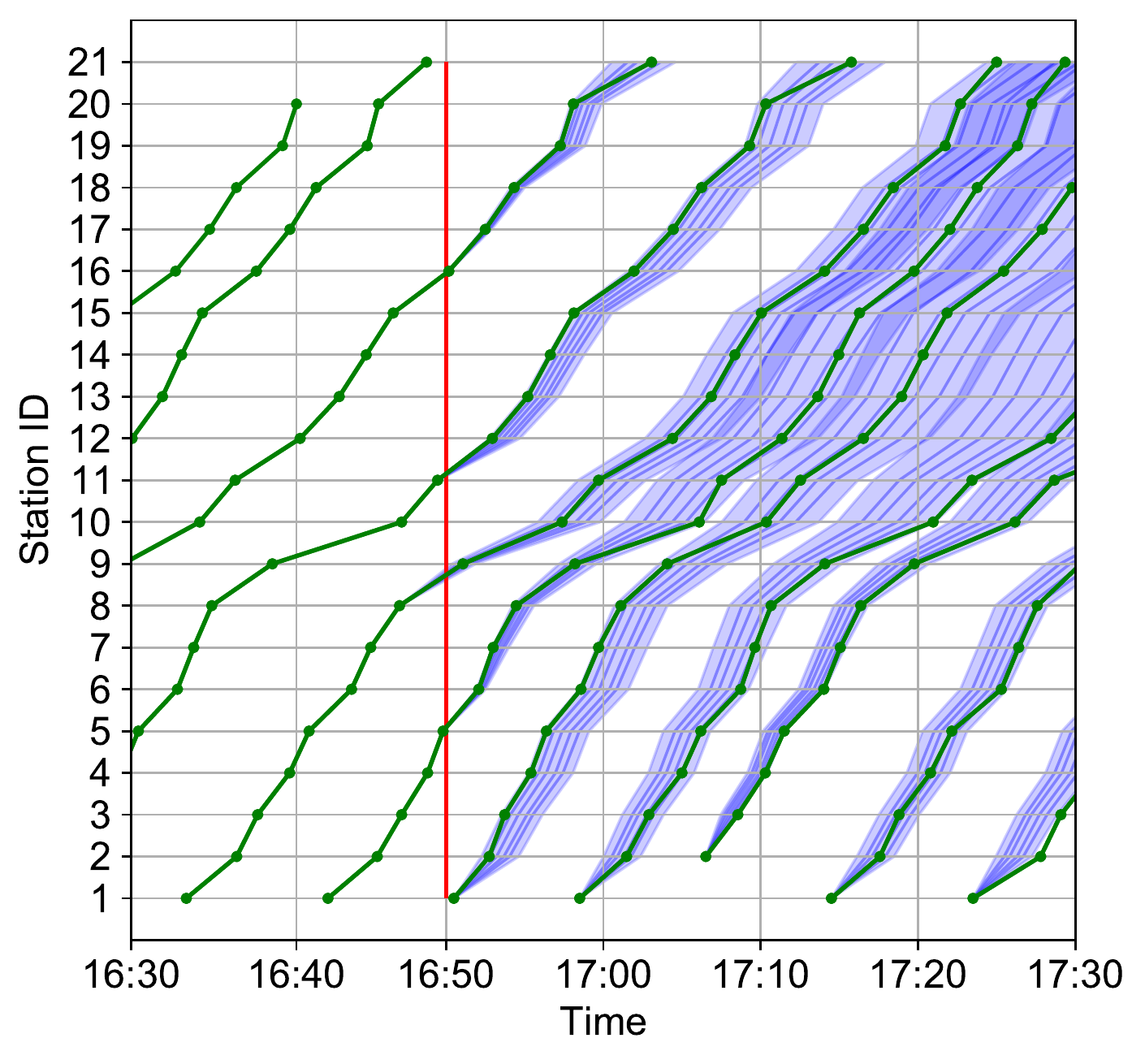}
}
\subfigure[Probabilistic forecasting for 17:05.]{
    \centering
    \includegraphics[width=0.47
    \textwidth]{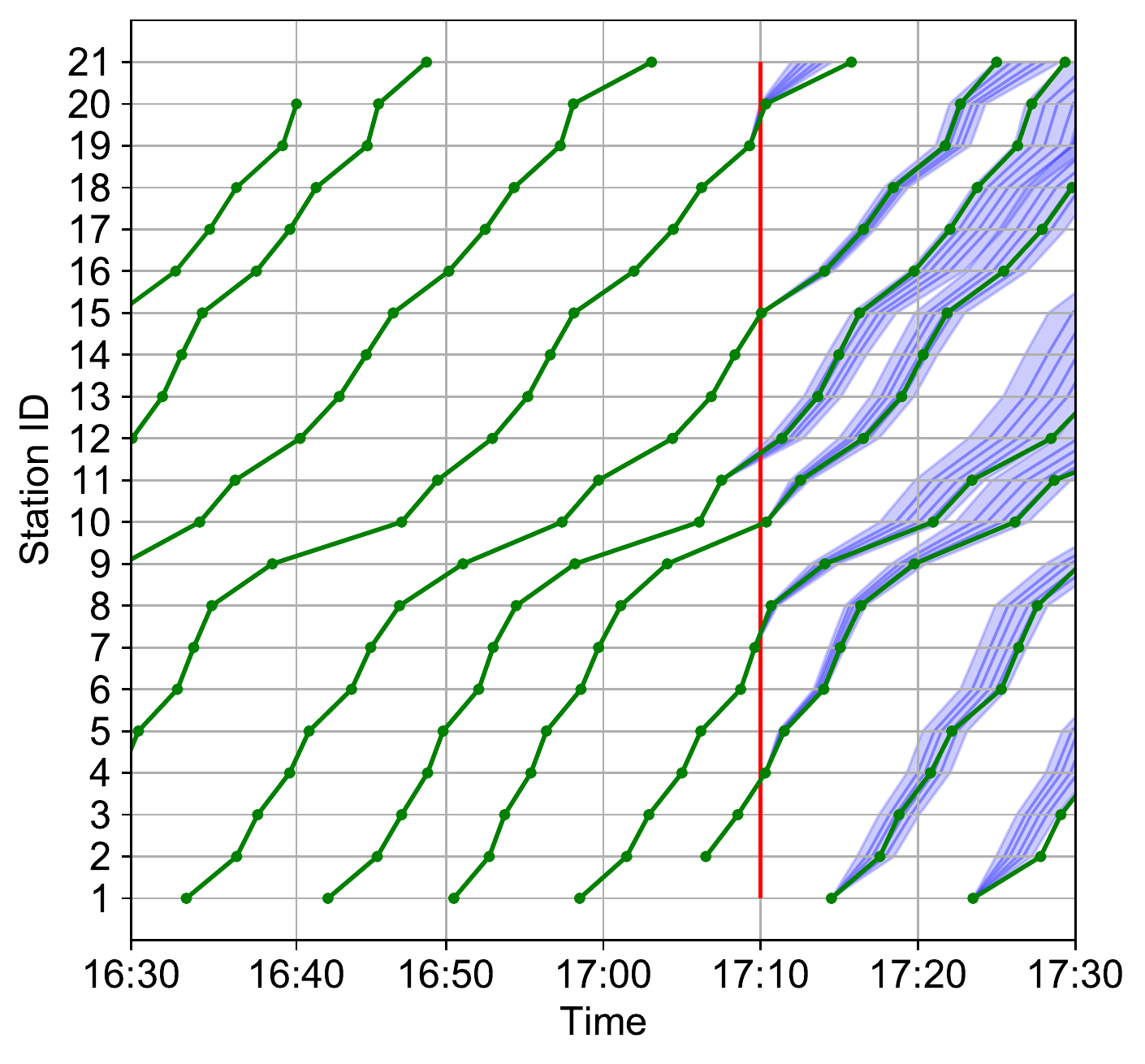}
}
\caption{Samples of probabilistic forecasting of route No.60.}
\label{fig:application}
\end{figure*}

We show the application of the probabilistic forecasting model for bus route No.60 of a selected day in Fig.~\ref{fig:application}. The left panel intuitively illustrates the probabilistic forecasting for bus travel time at 16:50 p.m., while the right panel corresponds to the forecasting at 17:10 p.m. All actual bus operations are shown by green lines in the time-space diagram. Assume that past operation of buses (i.e., before 16:50 p.m. and before 17:10 p.m.) are observed and we use the observed information to forecast the future time-space position of buses. The variability of the probabilistic forecasting is shown with the 10th, 25th, 40th, 60th, 75th, and 90th percentile values. We can see that the proposed model can make a good probabilistic forecasting for bus arrival time. By comparing the two figures, we can also find that for a bus trip, more observed links can help improve the forecasting accuracy and reduce the forecasting variability.

\begin{figure}[!ht]
\centering
\subfigure[Forecasting for observation 1.]{
    \centering
    \includegraphics[width = 0.95\textwidth]{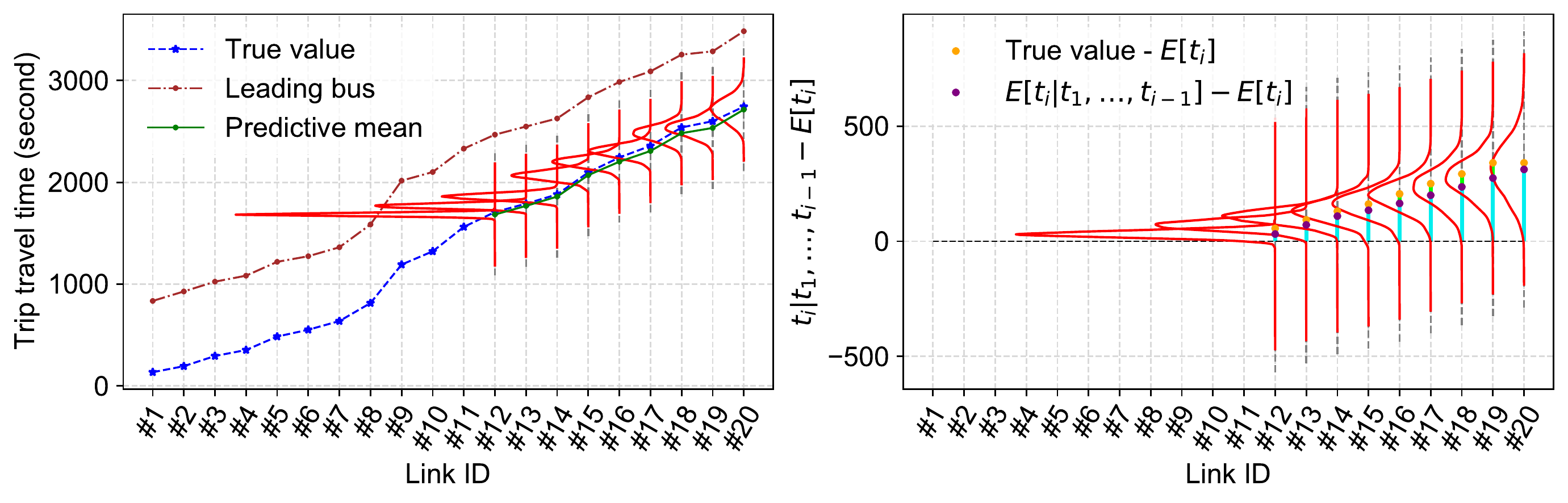}}
\subfigure[Forecasting for observation 2.]{
    \centering
    \includegraphics[width=0.95
    \textwidth]{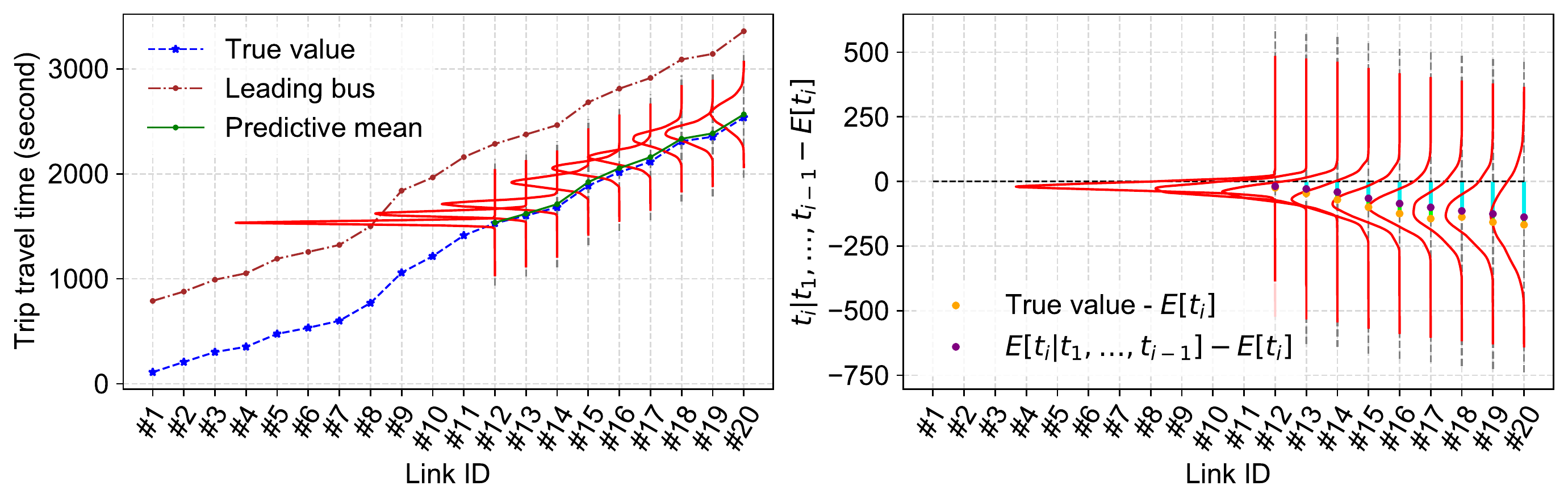}
}
\caption{The probabilistic forecasting for trip travel time.}
\label{probability_forecasting}
\end{figure}

Finally, we use the two bus trips to show the forecasting distributions. Fig.~\ref{probability_forecasting} shows the probabilistic forecasting results. In this experiment, we assume that we have observed the first eleven link travel times of the following bus and the corresponding link travel times/headways of the leading bus, then we intend to make probabilistic forecasting for the trip travel times of the last several links. In the left panel, the brown points are the trip travel times of the leading bus; the blue points are the true trip travel times of the following bus, and the green points are the predictive mean values. We can see that the predictive mean values fit the actual values quite well, demonstrating the proposed probabilistic model can achieve accurate forecasts. Moreover, the red bell curves are the trip travel time distributions, and we can find that the red bell curves are fatter with the increasing number of links in a trip, indicating the variance is increasing. The right panels show the mean corrected estimation, and the purple points (we refer to them as corrected mean values) are computed by posterior conditional mean values minus model mean values; the orange points are computed by the difference between true values and model mean values. We can find that the posterior conditional mean can make a more accurate prediction than the model mean. If we do not use the information of the observed link travel times/headways, the forecasting mean vectors should be the model mean vectors. As can be seen, the corrected mean values for observation 1 as shown in Fig.~\ref{probability_forecasting} (a) are larger than zero, while the corrected mean values for observation 2 as shown in Fig.~\ref{probability_forecasting} (b) are lower than zero, demonstrating the observed information indeed reinforces the forecasting for the upcoming links.

\section{Conclusion}
\label{conclusion}

In this paper, we propose a new representation for bus link travel time and model it with Multivariate Gaussian mixture distributions for probabilistic bus travel time forecasting. Our approach captures/handles the link travel time correlations of a bus route, the interactions between adjacent buses, the multimodality of bus travel time distribution, and missing values in data. We also integrate the Gaussian mixture model with a topic model framework to capture bus travel time patterns in different periods of a day. We conduct the experiments on a real-world dataset to evaluate the proposed model; results show the proposed model that considers the dependencies between adjacent buses and the headway relationships significantly outperforms baseline models that do not
consider these factors, in terms of both predictive means and distributions. Besides forecasting, the parameters of the proposed model reveal the difference between mixture patterns in terms of the time interval of a day, the mean vector, and the correlation matrix.

Besides making probabilistic forecasting, our approach has other potential practice and research implications. First, the parameters of the proposed model contain rich information for bus agencies to build better timetables and schedules and to evaluate the resilience and reliability of timetables/schedules. For example, analyzing link travel time correlations and delay propagation using correlation matrices, and understanding temporal patterns of the bus route from mixing coefficients. Second, with sufficient training data, the proposed model can also be used to have probabilistic simulations about train/bus operation and delay propagation. Third, our proposed framework can incorporate further information, such as passenger demand, onboard passengers, etc., to reinforce the probabilistic forecasting for bus travel time.

Our proposed probabilistic forecasting model for bus travel time only considers using one leading bus. In practice, several leading buses could be correlated with the following bus; therefore, we can incorporate more leading buses to reinforce the probabilistic travel time forecasting for the following bus. However, the increasing number of leading buses will result in a significant increase in the variable's dimension; the proposed Bayesian Gaussian mixture model is not suitable for high-dimensional variable vectors (e.g., $n>100$) as the covariance matrix will be large, and the computation is highly expensive. In this case, we will consider using a mixture of probabilistic principal component analysis (PCA) to reduce dimensionality. Our further research will utilize the mixture of probabilistic PCA to model the following bus with more leading buses to make probabilistic forecasting for bus travel time.

\section*{Acknowledgements}
This research is supported in part by the
Fonds de Recherche du Qu\'{e}bec-Soci\'{e}t\'{e} et Culture (FRQSC) under the NSFC-FRQSC Research Program on Smart Cities and Big Data, the Canada Foundation for Innovation (CFI) John R. Evans Leaders Fund, and the Natural Sciences and Engineering Research Council (NSERC) of Canada. X. Chen acknowledges funding support from the China Scholarship Council (CSC).

%
%
%

\bibliographystyle{elsarticle-harv}
\bibliography{references}

\pagebreak
\appendix


\counterwithin{figure}{section}

\section{Figure}

\begin{figure*}[!ht]
\centering

\subfigure{
    \centering
    \includegraphics[width=0.9
    \textwidth]{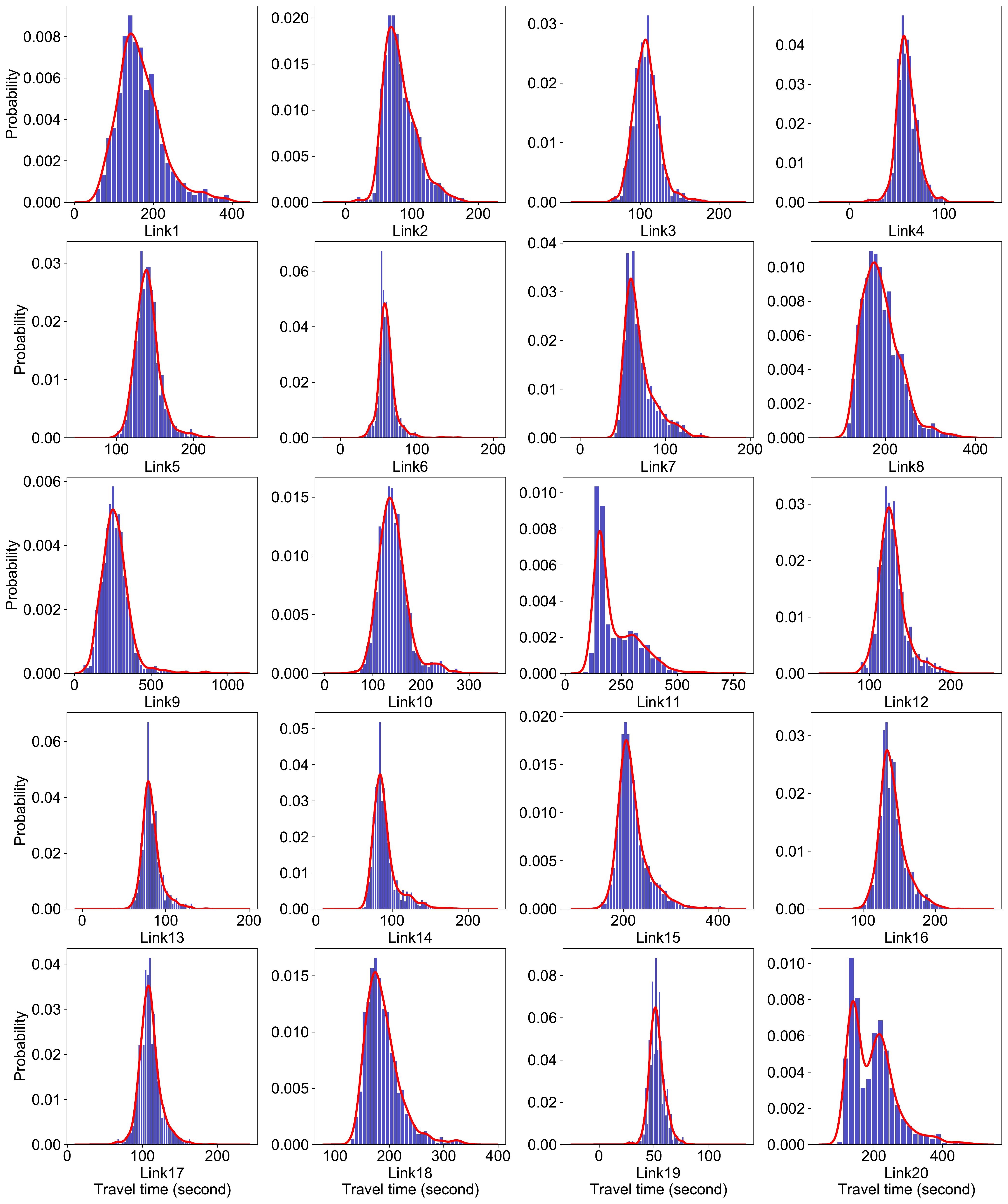}
}
\caption{Empirical distribution of link travel time of bus route No.60.}
\label{Empdis}
\end{figure*}


\end{document}